\documentclass[graybox]{svmult}

\usepackage{mathptmx}       
\usepackage{helvet}         
\usepackage{courier}        
\usepackage{type1cm}        
\usepackage{makeidx}         
\usepackage{graphicx}        
\usepackage{multicol}        
\usepackage[bottom]{footmisc}

\usepackage{amsmath,amssymb,amsfonts,amscd}
\usepackage{cite}

\newcommand{\bq}{\begin{eqnarray}}
\newcommand{\eq}{\end{eqnarray}}
\newcommand{\eps}{\varepsilon}

\usepackage[OT2,T1]{fontenc}
\DeclareSymbolFont{cyrletters}{OT2}{wncyr}{m}{n}
\DeclareMathSymbol{\Sha}{\mathalpha}{cyrletters}{"58}

\makeindex             

\begin{document}

\title*{Hopf algebras and Dyson-Schwinger equations}

\author{Stefan Weinzierl}
\institute{Stefan Weinzierl \at PRISMA Cluster of Excellence, Institut f{\"u}r Physik, Johannes Gutenberg-Universit{\"a}t Mainz, 
D - 55099 Mainz, Germany, \email{weinzierl@uni-mainz.de}}
%
%
\maketitle

\abstract{
In these lectures I discuss Hopf algebras and Dyson-Schwinger equations.
The lectures start with an introduction to Hopf algebras,
followed by a review of the contribution and application of Hopf algebras to particle physics.
The final part of these lectures is devoted to the relation between Hopf algebras and Dyson-Schwinger equations.
}


\section{Hopf algebras}
\label{sec:hopf}

Let us start with a brief history of Hopf algebras:
Hopf algebras were originally introduced to mathematics in 1941 to enable  
similar aspects of groups and algebras to be described in a unified manner \cite{Hopf}.
An article by Woronowicz in 1987 \cite{Woronowicz}, 
which provided explicit examples of non-trivial (non-co-commutative) Hopf algebras, 
triggered the interest of the physics community.
This led to applications of Hopf algebras in the field of integrable systems and quantum groups.
In physics, Hopf algebras received a further boost in 1998, when
Kreimer and Connes re-examined the renormalization of quantum field theories and showed
that they can be described by a Hopf algebra structure \cite{Kreimer:1998dp,Connes:1998qv}.
Since then, Hopf algebras have appeared in several facets of physics.

Let us now consider the definition of a Hopf algebra.
The presentation in this section closely follows that in \cite{Weinzierl:2003ub}.
References for further reading are \cite{Sweedler,Kassel,Majid:1990vz,Manchon:2004,Frabetti:2008}.
Let $R$ be a commutative ring with unit $1$.
An algebra over the ring $R$ is an $R$-module together with a multiplication $\cdot$ and a unit $e$.
We will always assume that the multiplication is associative.
In physics, the ring $R$ will almost always be a field $K$
(examples are the rational numbers ${\mathbb Q}$, the real numbers ${\mathbb R}$, or the complex number ${\mathbb C}$). 
In this case the $R$-module will actually be a $K$-vector space.
Note that the unit $e$ can be viewed as a map from $R$ to $A$ and that the multiplication $\cdot$
can be viewed as a map from the tensor product $A \otimes A$ to $A$ 
(e.g., one takes two elements from $A$, multiplies them, and obtains one element as the outcome):
\begin{align}
 & \mbox{Multiplication:} &  \cdot \; : & \;\; A \otimes A \rightarrow A,
 \nonumber \\
 & \mbox{Unit:} & e \; : & \;\; R \rightarrow A.
\end{align}
Instead of multiplication and a unit, a co-algebra has the dual structures:
a co-multiplication $\Delta$ and a co-unit $\bar{e}$.
The co-unit $\bar{e}$ is a map from $A$ to $R$, whereas the co-multiplication $\Delta$ is a map from $A$ to
$A \otimes A$:
\begin{align}
 & \mbox{Co-multiplication:} &  \Delta \; : & \;\; A \rightarrow A \otimes A, 
 \nonumber \\
 & \mbox{Counit:} & \bar{e} \; : & \;\; A \rightarrow R.
\end{align}
Note that mapping of a co-multiplication and co-unit proceeds in the reverse direction compared to multiplication
and unit.
We will always assume that the co-multiplication $\Delta$ is co-associative.
What does co-associativity mean? We can easily derive it from associativity as follows:
For $a,b,c \in A$ associativity requires
\bq
\label{condition_associativity}
 \left( a \cdot b \right) \cdot c
 & = & 
 a \cdot \left( b \cdot c \right).
\eq
We can re-write condition~(\ref{condition_associativity}) in the form of a commutative diagram:
\bq
\begin{CD}
A \otimes A \otimes A @>{\mathrm{id} \otimes \cdot}>> A \otimes A \\
@VV{\cdot \otimes \mathrm{id}}V @VV{\cdot}V \\
A \otimes A @>{\cdot}>> A \\
\end{CD}
\eq
We obtain the condition for co-associativity by reversing all arrows and by exchanging multiplication with
co-multiplication. We thus obtain the following commutative diagram:
\bq
\begin{CD}
A @>{\Delta}>> A \otimes A \\
@VV{\Delta}V @VV{\Delta \otimes \mathrm{id}}V \\
A \otimes A @>{\mathrm{id} \otimes \Delta}>> A \otimes A \otimes A  \\
\end{CD}
\eq
The general form of the co-product is
\bq
\Delta(a) & = & \sum\limits_i a_i^{(1)} \otimes a_i^{(2)},
\eq
where $a_i^{(1)}$ denotes an element of $A$ appearing 
in the first slot of $A \otimes A$ and $a_i^{(2)}$ correspondingly denotes an element of $A$ 
appearing in the second slot.
Sweedler's notation \cite{Sweedler} consists of omitting the dummy index $i$ and the summation symbol:
\bq
\Delta(a) & = & 
a^{(1)} \otimes a^{(2)}
\eq 
The sum is implicitly understood. 
This is similar to Einstein's summation convention, except that the dummy summation index $i$ is also dropped. 
The superscripts ${}^{(1)}$ and ${}^{(2)}$ indicate that a sum is involved.
Using Sweedler's notation, co-associativity is equivalent to
\bq
 a^{(1) (1)} \otimes a^{(1) (2)} \otimes a^{(2)}
 & = &
 a^{(1)} \otimes a^{(2) (1)} \otimes a^{(2) (2)}.
\eq
As it is irrelevant whether we exchange the second co-product with the first or the second factor in the tensor product,
we can simply write
\bq
 \Delta^2\left(a\right)
 & = &
 a^{(1)} \otimes a^{(2)} \otimes a^{(3)}.
\eq
If the co-product of an element $a \in A$ is of the form
\bq
 \Delta\left(a\right) 
 & = & 
 a \otimes a,
\eq
then $a$ is referred to as a group-like element.
If the co-product of $a$ is of the form
\bq
 \Delta\left(a\right)
 & = & 
 a \otimes e + e \otimes a,
\eq
then $a$ is referred to as a primitive element.

In an algebra we have for the unit $1$ of the underlying ring $R$ and the unit $e$ of the algebra
the relation
\bq
 a 
 \;\; = \;\;
 1 \cdot a
 \;\; = \;\;
 e \cdot a
 \;\; = \;\;
 a
\eq
for any element $a \in A$ (together with the analog relation $a = a \cdot 1 = a \cdot e = a$).
In terms of commutative diagrams this is expressed as
\bq
\label{axiom_unit}
\begin{CD}
A \otimes A @= A \otimes A \\
 @A{e \otimes \mathrm{id}}AA @VV{\cdot}V \\
R \otimes A @=^{\!\!\!\!\!\! \!\!\!\!\!\! \!\!\!\!\!\! \!\!\!\!\!\! \!\!\!\!\!\! \!\!\!\!\!\! \!\!\! \cong \,\,\,\,\,\, \,\,\,\,\,\, \,\,\,\,\,\, \,\,\,\,\,\, \,\,\,\,\,\,} A \\
\end{CD}
 & &
 \hspace*{15mm}
\begin{CD}
A \otimes A @= A \otimes A \\
 @A{\mathrm{id} \otimes e}AA @VV{\cdot}V \\
A \otimes R @=^{\!\!\!\!\!\! \!\!\!\!\!\! \!\!\!\!\!\! \!\!\!\!\!\! \!\!\!\!\!\! \!\!\!\!\!\! \!\!\! \cong \,\,\,\,\,\, \,\,\,\,\,\, \,\,\,\,\,\, \,\,\,\,\,\, \,\,\,\,\,\,} A \\
\end{CD}
\eq
In a co-algebra we have the dual relations obtained from eq.~(\ref{axiom_unit}) by reversing all arrows
and by exchanging multiplication with co-multiplication as well as by exchanging the unit $e$ with the co-unit $\bar{e}$:
\bq
\begin{CD}
A \otimes A @= A \otimes A \\
 @V{\bar{e} \otimes \mathrm{id}}VV @AA{\Delta}A \\
R \otimes A @=^{\!\!\!\!\!\! \!\!\!\!\!\! \!\!\!\!\!\! \!\!\!\!\!\! \!\!\!\!\!\! \!\!\!\!\!\! \!\!\! \cong \,\,\,\,\,\, \,\,\,\,\,\, \,\,\,\,\,\, \,\,\,\,\,\, \,\,\,\,\,\,} A \\
\end{CD}
 & &
 \hspace*{15mm}
\begin{CD}
A \otimes A @= A \otimes A \\
 @V{\mathrm{id} \otimes \bar{e}}VV @AA{\Delta}A \\
A \otimes R @=^{\!\!\!\!\!\! \!\!\!\!\!\! \!\!\!\!\!\! \!\!\!\!\!\! \!\!\!\!\!\! \!\!\!\!\!\! \!\!\! \cong \,\,\,\,\,\, \,\,\,\,\,\, \,\,\,\,\,\, \,\,\,\,\,\, \,\,\,\,\,\,} A \\
\end{CD}
\eq

A bi-algebra is an algebra and a co-algebra at the same time,
such that the two structures are compatible with each other.
In terms of commutative diagrams, the compatibility condition between the product and the
co-product is expressed as
\bq
\begin{CD}
A \otimes A @>{\cdot}>> A @>{\Delta}>> A \otimes A \\
@VV{\Delta \otimes \Delta}V & & @AA{\cdot \otimes \cdot}A \\
A \otimes A \otimes A \otimes A & @>{\mathrm{id} \otimes \tau \otimes \mathrm{id}}>> & A \otimes A \otimes A \otimes A\\
\end{CD}
\eq
where $\tau : A \otimes A \rightarrow A \otimes A$ is the map, which exchanges the entries in the two slots:
$\tau(a \otimes b) = b \otimes a$.
Using Sweedler's notation, the compatibility between the multiplication and co-multiplication is expressed as
\bq
\label{bialg}
 \Delta\left( a \cdot b \right)
 & = &
\left( a^{(1)} \cdot b^{(1)} \right)
 \otimes \left( a^{(2)} \cdot b^{(2)} \right).
\eq
It is common practice to write the right-hand side of eq.~(\ref{bialg}) as
\bq
\left( a^{(1)} \cdot b^{(1)} \right)
 \otimes \left( a^{(2)} \cdot b^{(2)} \right)
 & = & 
 \Delta\left(a\right) \Delta\left(b\right).
\eq
In addition, there is a compatibility condition between the unit and the co-product
\bq
\label{compatibility_unit_coproduct}
\begin{CD}
 R \otimes R \cong R @>{e}>> A \\
 @V{e \otimes e}VV @VV{\Delta}V \\
 A \otimes A @= A \otimes A \\
\end{CD}
\eq
as well as a compatibility condition between the co-unit and the product, which is dual to eq.~(\ref{compatibility_unit_coproduct}):
\bq
\label{compatibility_counit_product}
\begin{CD}
 A @>{\bar{e}}>> R \cong R \otimes R \\
 @A{\cdot}AA @AA{\bar{e} \otimes \bar{e}}A \\
 A \otimes A @= A \otimes A \\
\end{CD}
\eq
The commutative diagrams in eq.~(\ref{compatibility_unit_coproduct}) and eq.~(\ref{compatibility_counit_product}) are equivalent to
\bq
 \Delta e = e \otimes e,
 \;\;\; \mbox{and} \;\;\; 
 \bar{e}\left(a \cdot b \right) = \bar{e}\left(a\right) \bar{e}\left(b\right),
 \;\;\; \mbox{respectively.}
\eq
A Hopf algebra is a bi-algebra with an additional map from $A$ to $A$, known as the 
antipode $S$, which fulfills
\bq
\label{antipode_def1}
\begin{CD}
A @>{\bar{e}}>> R @>{e}>> A \\
@VV{\Delta}V & & @AA{\cdot}A \\
A \otimes A & @>{\mathrm{id} \otimes S}>{S \otimes \mathrm{id}}> & A \otimes A\\
\end{CD}
\eq
An equivalent formulation is
\bq
\label{antipode_def2}
 a^{(1)} \cdot S\left( a^{(2)} \right)
 \;\; = \;\;
 S\left(a^{(1)}\right) \cdot a^{(2)} 
 \;\; = \;\;
 e \cdot \bar{e}(a).
\eq
A bi-algebra that has an antipode (satisfying the commutative diagram~(\ref{antipode_def1}) 
or eq.~(\ref{antipode_def2})) is unique.

If a Hopf algebra $A$ is either commutative or co-commutative, then
\bq
 S^2 & = & \mathrm{id}.
\eq
A bi-algebra $A$ is graded, if it has a decomposition
\bq
 A & = &
 \bigoplus\limits_{n \ge 0} A_n,
\eq
with
\bq
 A_n \cdot A_m \subseteq A_{n+m},
 & &
 \;\;\;
 \Delta\left(A_n\right) \subseteq \bigoplus\limits_{k+l=n} A_k \otimes A_l.
\eq
Elements in $A_n$ are said to have degree $n$.
The bi-algebra is graded connected, if in addition one has
\bq
 A_0 & = & R \cdot e.
\eq
It is useful to know that a graded connected bi-algebra is automatically a Hopf algebra \cite{Ehrenborg}.

An algebra $A$ is commutative if for all $a,b \in A$ one has
\bq
\label{def_commutative}
 a \cdot b & = & b \cdot a.
\eq
A co-algebra $A$ is co-commutative if for all $a \in A$ one has
\bq
\label{def_cocommutative}
 a^{(1)} \otimes a^{(2)} 
 & = &
 a^{(2)} \otimes a^{(1)}.
\eq
With the help of the swap map $\tau$ we may express commutativity and co-commutativity equivalently as
\bq
 \cdot \tau = \cdot,
 \;\;\; \mbox{and} \;\;\;
 \tau \Delta = \Delta,
 \;\;\; \mbox{respectively.}
\eq 
Let us now consider a few examples of Hopf algebras.
\begin{enumerate}

\item The group algebra.
Let $G$ be a group and denote by $KG$ the vector space with basis $G$ over the field $K$.
Then $KG$ is an algebra with the multiplication given by the group multiplication. 
The co-unit, the co-product, and the antipode are defined for the basis elements $g \in G$ as follows: 
The co-unit $\bar{e}$ is given by:
\bq
 \bar{e}\left( g\right) 
 & = & 
 1.
\eq
The co-product $\Delta$ is given by:
\bq
 \Delta\left( g\right) 
 & = & 
 g \otimes g.
\eq
The antipode $S$ is given by:
\bq
 S\left( g \right) 
 & = & 
 g^{-1}.
\eq
Having defined the co-unit, the co-product, and the antipode for the basis elements $g \in G$, 
the corresponding definitions for arbitrary vectors in $KG$ 
are obtained by linear extension.
$KG$ is a co-commutative Hopf algebra, which means that $KG$ is commutative if $G$ is commutative.

\item Lie algebras.
A Lie algebra ${\mathfrak g}$ is not necessarily associative nor does it have a unit.
To overcome this obstacle one considers the universal enveloping algebra $U({\mathfrak g})$,
obtained from the tensor algebra $T({\mathfrak g})$ by factoring out the ideal generated by
\bq
 X \otimes Y - Y \otimes X - \left[ X, Y \right],
\eq
with $X, Y \in {\mathfrak g}$.
The co-unit $\bar{e}$ is given by:
\bq
 \bar{e}\left( e\right) = 1,
 & &
 \bar{e}\left( X\right) = 0.
\eq
The co-product $\Delta$ is given by:
\bq
 \Delta(e) = e \otimes e, 
 & &
 \Delta(X) = X \otimes e + e \otimes X.
\eq
The antipode $S$ is given by:
\bq
 S(e) = e, 
 & &
 S(X) = -X.
\eq

\item Quantum SU(2).
The Lie algebra $su(2)$ is generated by three generators $H$, $X_\pm$ with
\bq
 \left[ H, X_\pm \right] = \pm 2 X_\pm,
 &  & 
 \left[ X_+, X_- \right] = H. 
\eq
To obtain the deformed algebra $U_q(su(2))$, the last relation is replaced with \cite{Majid:1990vz,Schupp:1993hn}
\bq
 \left[ X_+, X_- \right] 
 & = & 
 \frac{q^H - q^{-H}}{q-q^{-1}}.
\eq
The undeformed Lie algebra $su(2)$ is recovered in the limit $q \rightarrow 1$. 
The co-unit $\bar{e}$ is given by:
\bq
 \bar{e}\left( e\right) = 1,
 & &
 \bar{e}\left( H\right) = \bar{e}\left( X_\pm\right) = 0.
\eq
The co-product $\Delta$ is given by:
\bq
 \Delta(H) 
 & = & 
 H \otimes e + e \otimes H, 
 \nonumber \\
 \Delta(X_\pm) 
 & = & 
 X_\pm \otimes q^{H/2} + q^{-H/2} \otimes X_\pm.
\eq
The antipode $S$ is given by:
\bq
 S(H) = -H, 
 & &
 S(X_\pm) = - q^{\pm 1} X_\pm .
\eq

\item Symmetric algebras.
Let $V$ be a finite dimensional vector space with basis $\{v_i\}$.
The symmetric algebra $S(V)$ is the direct sum
\bq
 S(V) 
 & = & 
 \bigoplus\limits_{n=0}^\infty S^n(V),
\eq
where $S^n(V)$ is spanned by elements of the form $v_{i_1} v_{i_2} ... v_{i_n}$ with
$i_1 \le i_2 \le ... \le i_n$.
The multiplication is defined by
\bq
 \left( v_{i_1} v_{i_2} ... v_{i_m} \right) \cdot \left( v_{i_{m+1}} v_{i_{m+2}} ... v_{i_{m+n}} \right)
 & = &
 v_{i_{\sigma(1)}} v_{i_{\sigma(2)}} ... v_{i_{\sigma(m+n)}},
\eq
where $\sigma$ is a permutation on $m+n$ elements such that $i_{\sigma(1)} \le i_{\sigma(2)} \le ... \le i_{\sigma(m+n)}$.
The co-unit $\bar{e}$ is given by:
\bq
 \bar{e}\left( e\right) = 1, \;\;\;
 & &
 \bar{e}\left( v_1 v_2 ... v_n\right) = 0.
\eq
The co-product $\Delta$ is given for the basis elements $v_i$ by:
\bq
 \Delta(v_i) 
 & = & 
 v_i \otimes e + e \otimes v_i.
\eq
Using (\ref{bialg})  one obtains for a general element of $S(V)$
\bq
 \Delta\left( v_1 v_2 ... v_n \right)
 & = &  
  v_1 v_2 ... v_n \otimes e
 + e \otimes v_1 v_2 ... v_n
 \nonumber \\
 & &
 + \sum\limits_{j=1}^{n-1} \sum\limits_\sigma
  v_{\sigma(1)} ... v_{\sigma(j)}
   \otimes 
  v_{\sigma(j+1)} ... v_{\sigma(n)},
\eq
where $\sigma$ runs over all $(j,n-j)$-shuffles. 
A $(j,n-j)$-shuffle is a permutation $\sigma$ of $(1,...,n)$ such that
\bq
 \sigma(1) < \sigma(2) < ... < \sigma(j)
 & \mbox{and} &
 \sigma(k+1) < ... < \sigma(n). \nonumber 
\eq
The antipode $S$ is given by:
\bq
 S( v_{i_1} v_{i_2} ... v_{i_n}) & = & (-1)^n v_{i_1} v_{i_2} ... v_{i_n}.
\eq

\item Shuffle algebras.
Consider a set of letters $A$. 
The set $A$ is known as the alphabet.
A word is an ordered sequence of letters:
\bq
 w & = & l_1 l_2 ... l_k,
\eq
where $l_1, ..., l_k \in A$.
The word of length zero is denoted by $e$.
The shuffle algebra ${\cal A}$ on the vector space spanned by words is defined by
\bq
 \left( l_1 l_2 ... l_k \right) \cdot \left( l_{k+1} ... l_r \right) 
 & = &
 \sum\limits_{\mathrm{shuffles} \; \sigma} l_{\sigma(1)} l_{\sigma(2)} ... l_{\sigma(r)},
\eq
where the sum runs over all permutations $\sigma$, which preserve the relative order
of $1,2,...,k$ and of $k+1,...,r$.
The name ``shuffle algebra'' is related to the analogy of shuffling cards: If a deck of cards
is divided into two parts and then shuffled, the relative order within the two individual parts
is conserved. A shuffle algebra is also known under the name ``mould symmetral'' \cite{Ecalle}.
The empty word $e$ is the unit in this algebra:
\bq
 e \cdot w = w \cdot e = w.
\eq
The recursive definition of the shuffle product is given by
\bq
\label{def_recursive_shuffle}
\left( l_1 l_2 ... l_k \right) \cdot \left( l_{k+1} ... l_r \right) = 
 l_1 \left[ \left( l_2 ... l_k \right) \cdot \left( l_{k+1} ... l_r \right) \right]
+
 l_{k+1} \left[ \left( l_1 l_2 ... l_k \right) \cdot \left( l_{k+2} ... l_r \right) \right].
 \nonumber \\
\eq
It is a well-known fact that the shuffle algebra is actually a (non-co-commutative) Hopf algebra \cite{Reutenauer}.
The co-unit $\bar{e}$ is given by:
\bq
 \bar{e}\left( e\right) = 1, \;\;\;
 & &
 \bar{e}\left( l_1 l_2 ... l_n\right) = 0.
\eq
The co-product $\Delta$ is given by:
\bq
 \Delta\left( l_1 l_2 ... l_k \right) 
 & = & 
 \sum\limits_{j=0}^k \left( l_{j+1} ... l_k \right) \otimes \left( l_1 ... l_j \right).
\eq
This particular co-product is also known as the deconcatenation co-product.
The antipode $S$ is given by:
\bq
 S\left( l_1 l_2 ... l_k \right) & = & (-1)^k \; l_k l_{k-1} ... l_2 l_1.
\eq
The shuffle algebra is generated by the Lyndon words \cite{Reutenauer}.
If one introduces a lexicographic ordering on the letters of the alphabet
$A$, a Lyndon word is defined by the property $w < v$
for any sub-words $u$ and $v$ such that $w= u v$.

\item Rooted trees. An individual rooted tree is shown in Fig.~(\ref{fig15}).
\begin{figure}
\begin{center}
\includegraphics[scale=0.8]{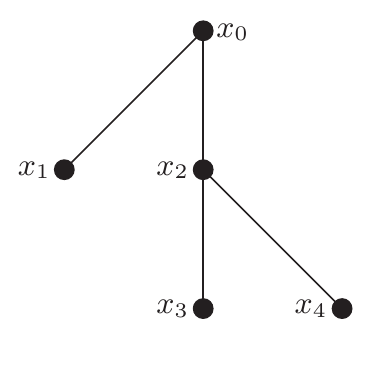}
\caption{\label{fig15} Illustration of a rooted tree. The root is drawn at the top and is labeled $x_0$.}
\end{center}
\end{figure}
We consider the algebra generated by rooted trees. Elements of this algebra are sets of rooted trees,
conventionally known as forests.
The product of two forests is simply the disjoint union of all trees from the two forests.
The empty forest, consisting of no trees, will be denoted by $e$.
Before we are able to define a co-product, we first need the definition of an admissible cut.
A single cut is a cut of an edge.
An admissible cut of a rooted tree is any assignment of single cuts such that any path from any vertex of the tree 
to the root has at most one single cut.
An admissible cut $C$ maps a tree $t$ to a monomial in trees $t_1 \cdot ... \cdot t_{n+1}$. 
Precisely one of these sub-trees $t_j$ will contain the root of $t$. 
We denote this distinguished tree by $R^C(t)$, and the monomial consisting of the $n$ other factors
by $P^C(t)$. 
The co-unit $\bar{e}$ is given by:
\bq
 \bar{e}(e) = 1, \;\;\;
 & &
 \bar{e}\left(t_1 \cdot ... \cdot t_k \right) = 0 \;\;\;\mbox{for}\; k \ge 1.
\eq
The co-product $\Delta$ is given by:
\bq
 \Delta(e) 
 & = & 
 e \otimes e, 
 \nonumber \\
 \Delta(t) 
 & = & 
 t \otimes e + e \otimes t + \sum\limits_{\mathrm{adm. cuts} \; C \mathrm{of} \; t} P^C(t) \otimes R^C(t),
 \nonumber \\
 \Delta\left(t_1 \cdot ... \cdot t_k\right)
 & = &
 \Delta\left(t_1\right) \; ... \; \Delta\left(t_k\right).
\eq
The antipode $S$ is given by:
\bq
 S(e) & = & e, 
 \nonumber \\
 S(t) 
 & = & 
 -t - \sum\limits_{\mathrm{adm. cuts} \; C \; \mathrm{of} \; t} S\left( P^C(t) \right) \cdot R^C(t),
 \nonumber \\
 S\left( t_1 \cdot ... \cdot t_k \right)
 & = &
 S\left(t_1\right) \cdot ... \cdot S\left(t_k\right).
\eq

\end{enumerate}
It is possible to classify the examples discussed above into four groups according to whether they are commutative or co-commutative.
\begin{itemize}
\item Commutative and co-commutative: 
Examples are the group algebra of a commutative group or the symmetric algebras.
\item Non-commutative and co-commutative: 
Examples are the group algebra of a non-commutative group or the universal enveloping algebra of a Lie algebra.
\item Commutative and non-co-commutative: 
Examples are shuffle algebra or the algebra of rooted trees.
\item Non-commutative and non-co-commutative: 
Examples are given by quantum groups.
\end{itemize}
Whereas research on quantum groups focused primarily on non-commutative and non-co-commutative Hopf algebras,
it turns out that for applications in perturbative quantum field theories commutative,
but not necessarily co-commutative, Hopf algebras such as shuffle algebras, symmetric algebras, and rooted trees 
are the most important.


\section{Applications in particle physics}
\label{sec:hopf_physics}

Let us now discuss two important applications of Hopf algebras in perturbative quantum field theory:
The combinatorics of renormalization and the Hopf algebras related to multiple polylogarithms.
The former topic is related to ultraviolet divergences occurring in Feynman integrals, whereas 
the latter topic concerns functions to which Feynman integrals evaluate.
The presentation in this section follows \cite{Weinzierl:2013yn}.
We start our discussion with a short introduction to Feynman integrals.

\subsection{Feynman integrals}
\label{subsection:feynman}

\begin{figure}
\begin{center}
\includegraphics[scale=0.8]{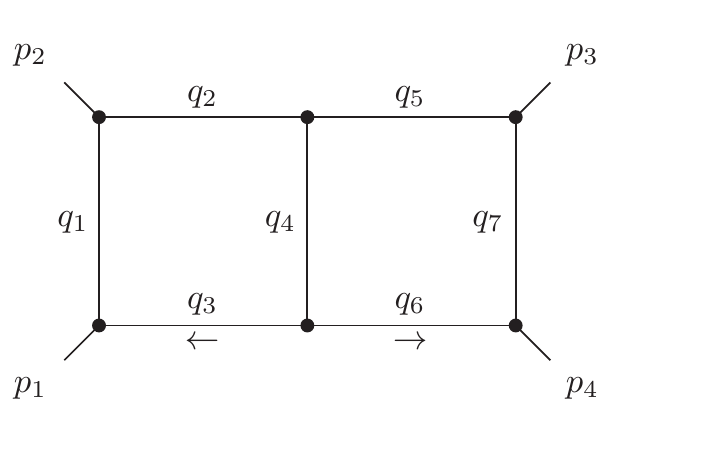}
\end{center}
\caption{\label{fig16} Illustration of a ``double box''-graph:
A two-loop Feynman diagram with four external and seven internal lines.
The momenta flowing out along the external lines and those flowing through the 
internal lines are labelled $p_1$, ..., $p_4$ and $q_1$, ..., $q_7$, respectively.}
\end{figure}

The perturbative expansion of quantum field theory can be organized in terms of Feynman graphs.
Feynman graphs can be considered as a pictorial notation for mathematical expressions 
arising in the context of perturbative quantum field theory. 
Fig.~(\ref{fig16}) shows an example of a Feynman graph.
Each part in a Feynman graph corresponds to a specific expression 
and the full Feynman graph corresponds to the product of these expressions.
For scalar theories the correspondence is as follows:
An internal edge corresponds to a propagator
\bq
 \frac{i}{q^2-m^2},
\eq
an external edge to the factor $1$. In scalar theories, a vertex also corresponds to the factor $1$.
In addition, for each internal momentum not constrained by momentum conservation, there is an integration
\bq 
 \int \frac{d^Dk}{\left(2\pi\right)^D},
\eq
where $D$ denotes the dimension of space-time.
Let us now consider a Feynman graph $G$ with $m$ external edges, $n$ internal edges, and $l$ loops.
With each internal edge we associate, apart from its momentum and its mass, a positive integer $\nu$, 
which provides the power to which the propagator occurs.
(We may think of $\nu$ as the relict of neglecting vertices of valency $2$. 
The number $\nu>1$ corresponds to $\nu-1$ mass insertions on this edge).
The momenta flowing through the internal lines can be expressed through the independent loop momenta
$k_1$, ..., $k_l$ and the external momenta $p_1$, ..., $p_m$ as 
\bq
 q_i & = & \sum\limits_{j=1}^l \rho_{ij} k_j + \sum\limits_{j=1}^m \sigma_{ij} p_j,
 \;\;\;\;\;\; 
 \rho_{ij}, \sigma_{ij} \in \{-1,0,1\}.
\eq
We define the Feynman integral by
\bq
\label{Feynman_integral_1}
I_G  & = &
 \left( \mu^2 \right)^{\nu-l D/2}
 \int \prod\limits_{r=1}^{l} \frac{d^Dk_r}{i\pi^{\frac{D}{2}}}\;
 \prod\limits_{j=1}^{n} \frac{1}{(-q_j^2+m_j^2)^{\nu_j}},
\eq
with $\nu=\nu_1+...+\nu_n$.
The inclusion of an arbitrary scale $\mu$, the factors $i \pi^{D/2}$ in the measure, and a minus sign for each propagator
are the conventions used in these lectures.
Feynman parametrization makes use of the identity
\bq
\label{feynman_parametrisation}
 \prod\limits_{j=1}^{n} \frac{1}{P_{j}^{\nu_j}} 
 & = &
 \frac{\Gamma\left(\nu\right)}{\prod\limits_{i=1}^n \Gamma\left(\nu_j\right)}
 \int\limits_\Delta \omega
 \left( \prod\limits_{i=1}^n x_i^{\nu_i-1} \right)
 \left( \sum\limits_{j=1}^{n} x_{j} P_{j} \right)^{-\nu},
\eq
where $\omega$ is a differential of the form $(n-1)$ given by
\bq
 \omega & = & \sum\limits_{j=1}^n (-1)^{j-1}
  \; x_j \; dx_1 \wedge ... \wedge \widehat{dx_j} \wedge ... \wedge dx_n.
\eq
The hat indicates that the corresponding term is omitted.
The integration is over
\bq
 \Delta & = & \left\{ \left[ x_1 : x_2 : ... : x_n \right] \in {\mathbb P}^{n-1} | x_i \ge 0, 1 \le i \le n \right\}.
\eq
We use eq.~(\ref{feynman_parametrisation}) with $P_j=-q_j^2+m_j^2$.
We can write
\bq
\label{eq_poly_calc_1}
 \sum\limits_{j=1}^{n} x_{j} (-q_j^2+m_j^2)
 & = & 
 - \sum\limits_{r=1}^{l} \sum\limits_{s=1}^{l} k_r M_{rs} k_s + \sum\limits_{r=1}^{l} 2 k_r \cdot Q_r + J,
\eq
where $M$ is a $l \times l$ matrix with scalar entries and $Q$ is a $l$-vector
with $D$-vectors as entries.
After Feynman parametrization, it becomes possible to integrate over the loop momenta $k_1$, ..., $k_l$ and we obtain
\bq
\label{Feynman_integral_2}
I_G  & = &
 \frac{\Gamma(\nu-lD/2)}{\prod\limits_{j=1}^{n}\Gamma(\nu_j)}
 \int\limits_{\Delta}  \omega
 \left( \prod\limits_{j=1}^n x_j^{\nu_j-1} \right)
 \frac{U^{\nu-(l+1) D/2}}{F^{\nu-l D/2}}.
\eq
The functions $U$ and $F$ are given by
\bq
\label{eq_poly_calc_2}
 U = \mbox{det}(M),
 & &
 F = \mbox{det}(M) \left( J + Q M^{-1} Q \right)/\mu^2,
\eq 
where $U$ and $F$ are both graph polynomials and have an alternative definition in terms of spanning trees and spanning forests \cite{Bogner:2010kv}.
A few remarks are in order:
The integral over the Feynman parameters is an $(n-1)$-dimensional integral in projective space ${\mathbb P}^{n-1}$, where $n$ is the number of internal edges of the graph.
Singularities may arise if the zero sets of $U$ and $F$ intersect the region of integration.
The dimension $D$ of space-time only appears in the exponents of the integrand and the exponents act as a regularization.
A Feynman integral has an expansion as a Laurent series in the parameter $\varepsilon=(4-D)/2$ of dimensional regularization:
\bq
\label{Feynman_integral_3}
 I_G & = &
 \sum\limits_{j=-2l}^\infty c_j \varepsilon^j.
\eq
The Laurent series of an $l$-loop integral can have poles in $\varepsilon$ up to the order $(2l)$. The poles in $\varepsilon$ correspond to
ultraviolet or infrared divergences.
The coefficients $c_j$ are functions of the scalar products $p_j \cdot p_k$, the masses $m_i$, and (in a trivial way) of the arbitrary scale $\mu$.

Transforming a Feynman integral from the form in eq.~(\ref{Feynman_integral_1}) to the form of eq.~(\ref{Feynman_integral_2})
is straightforward and will be illustrated by an example below.
The challenging part is to obtain the expansion in eq.~(\ref{Feynman_integral_3}) and to find explicit expressions for the coefficients $c_j$
in eq.~(\ref{Feynman_integral_3}).

As an example for the transition from eq.~(\ref{Feynman_integral_1}) to eq.~(\ref{Feynman_integral_2})
let us consider the two-loop double box graph in fig.~(\ref{fig16}).
In fig.~(\ref{fig16}) there are two independent loop momenta. 
We can choose them to be $k_1=q_3$ and $k_2=q_6$.
Then all other internal momenta are expressed in terms of $k_1$, $k_2$, 
and the external momenta $p_1$, ..., $p_4$:
\bq
\begin{array}{lll}
 q_1 = k_1 - p_1,
&
 q_2 = k_1 - p_1 - p_2,
&
 q_4 = k_1 + k_2,
 \\
 q_5 = k_2 - p_3 - p_4,
 &
 q_7 = k_2 - p_4.
 & \\
\end{array}
\eq
We will consider the case
\bq
\label{specification_double_box}
 & & p_1^2 = 0, \;\;\; p_2^2 = 0, \;\;\; p_3^2 = 0, \;\;\; p_4^2 = 0,
 \nonumber \\
 & & m_1 = m_2 = m_3 = m_4 = m_5 = m_6 = m_7 = 0.
\eq
We define
\bq
 s = \left(p_1+p_2\right)^2=\left(p_3+p_4\right)^2,
 & &
 t = \left(p_2+p_3\right)^2=\left(p_1+p_4\right)^2.
\eq
We have
\bq
\lefteqn{
 \sum\limits_{j=1}^7 x_j \left(-q_j^2\right) 
 =
 - \left(x_1+x_2+x_3+x_4\right) k_1^2 - 2 x_4 k_1 \cdot k_2 - \left( x_4+x_5+x_6+x_7\right) k_2^2
} & & \nonumber \\
 & &
 + 2 \left[ x_1 p_1 + x_2 \left( p_1 + p_2 \right) \right] \cdot k_1
 + 2 \left[ x_5 \left( p_3 + p_4 \right) + x_7 p_4 \right] \cdot k_2
 - \left( x_2 + x_5 \right) s.
\eq
In comparison with eq.~(\ref{eq_poly_calc_1})
we find
\bq
 M & = & \left( \begin{array}{cc}
 x_1+x_2+x_3+x_4 & x_4 \\
 x_4 & x_4+x_5+x_6+x_7 \\
 \end{array} \right),
 \nonumber \\
 Q & = & \left( \begin{array}{c}
          x_1 p_1 + x_2 \left( p_1 + p_2 \right) \\
          x_5 \left( p_3 + p_4 \right) + x_7 p_4 \\
          \end{array} \right),
 \nonumber \\
 J & = & \left( x_2 + x_5 \right) \left(-s\right).
\eq
Plugging this into eq.~(\ref{eq_poly_calc_2})
we obtain the graph polynomials as
\bq
 U & = & \left( x_1+x_2+x_3 \right) \left( x_5+x_6+x_7 \right) + x_4 \left( x_1+x_2+x_3+x_5+x_6+x_7 \right),
 \nonumber \\
 F & = & \left[ x_2 x_3 \left( x_4+x_5+x_6+x_7 \right)
                        + x_5 x_6 \left( x_1+x_2+x_3+x_4 \right)
                        + x_2 x_4 x_6 + x_3 x_4 x_5 \right] \left( \frac{-s}{\mu^2} \right)
 \nonumber \\
 & &
      + x_1 x_4 x_7 \left( \frac{-t}{\mu^2} \right).
\eq

\subsection{Renormalization}
\label{subsect:renorm}

\begin{figure}
\begin{center}
\includegraphics[scale=0.8]{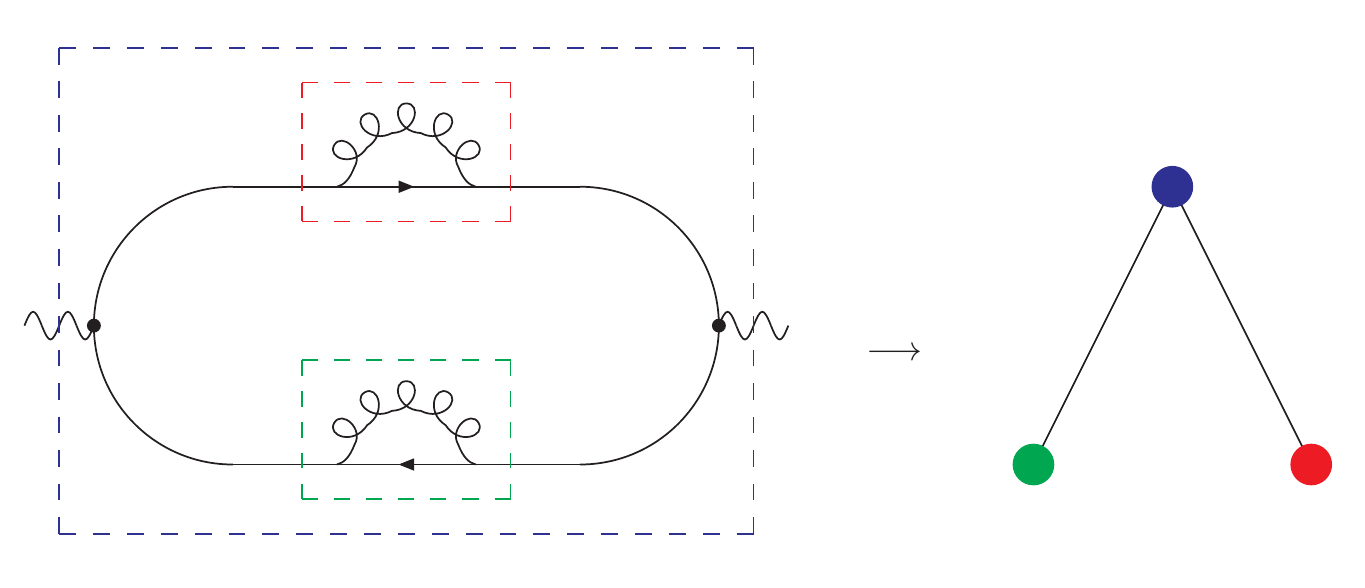}
\caption{\label{fig1} Three-loop two-point function with an overall ultraviolet divergence and two sub-divergences.
We find the corresponding rooted tree by first drawing boxes around all ultraviolet-divergent sub-graphs.
The rooted tree is obtained from the nested structure of these boxes.}
\end{center}
\end{figure}

Let us now consider the ultraviolet (or short-distance) singularities of Feynman integrals.
These singularities are removed by renormalization \cite{Zimmermann:1969jj}.
The combinatorics involved in the renormalization are governed by a Hopf algebra \cite{Kreimer:1998dp,Connes:1998qv}.
The relevant Hopf algebra is that which is generated by rooted trees.
We determine the relation between a Feynman graph and the corresponding rooted trees by starting from the fact
that sub-graphs may give rise to sub-divergences.
That is, the rooted trees encode the nested structure of sub-divergences.
This is best explained by an example.
Fig.~(\ref{fig1}) shows a three-loop two-point function. This Feynman integral has an overall ultraviolet divergence
and two sub-divergences, corresponding to the two fermion self-energy corrections.
We obtain the corresponding rooted tree by drawing boxes around all ultraviolet-divergent sub-graphs.
The rooted tree is obtained from the nested structure of these boxes.
\begin{figure}
\begin{center}
\includegraphics[scale=0.8]{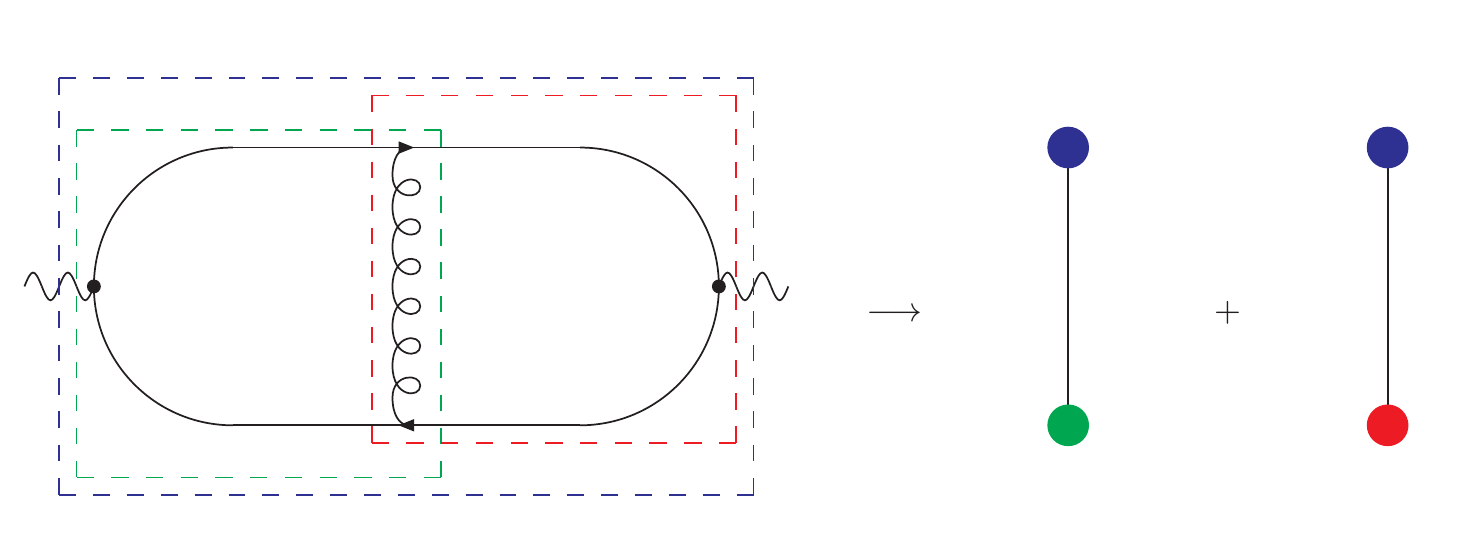}
\caption{\label{fig2} Example with overlapping singularities. This graph corresponds to a sum of rooted trees}
\end{center}
\end{figure}
Graphs with overlapping singularities correspond to a sum of rooted trees. This is illustrated for a two-loop example 
with an overlapping singularity in fig.~(\ref{fig2}).

We recall that the co-unit applied to any non-trivial rooted tree $t\neq e$ yields zero:
\bq
 \bar{e}\left(t\right) & = & 0,
 \;\;\;\;\;\; t \neq e.
\eq
Let us further recall the recursive definition of the antipode for the Hopf algebra of rooted trees:
\bq
 S(t) & = & -t - \sum\limits_{\mathrm{adm. cuts} \; C \; \mathrm{of} \; t} S\left( P^C(t) \right) \cdot R^C(t).
\eq
The antipode satisfies (see eq.~(\ref{antipode_def2})) for any non-trivial rooted tree $t \neq e$
\bq
\label{untwisted}
 S\left(t^{(1)}\right) t^{(2)} & = & 0,
\eq
where we used Sweedler's notation.
Eq.~(\ref{untwisted}) will be our starting point.
However, rather than obtaining zero on the right-hand side,
we are interested in a finite quantity.
This can be achieved as follows:
Let $R$ be an operation, which approximates a tree by another tree with the same singularity structure,
and which satisfies the Rota-Baxter relation \cite{EbrahimiFard:2006iy}:
\bq
\label{rotabaxter}
 R\left( t_1 t_2 \right) + R\left( t_1 \right) R\left( t_2 \right) 
 & = &
 R\left( t_1 R\left( t_2 \right) \right) + R\left( R\left( t_1 \right) t_2 \right).
\eq
In physics, we may think about $R(t)$ as the appropriate counter-terms.
For example, minimal subtraction ($\overline{MS}$) 
\bq
 R\left( \sum\limits_{k=-L}^\infty c_k \eps^k \right) 
 & = & 
 \sum\limits_{k=-L}^{-1} c_k \eps^k
\eq
fulfills the Rota-Baxter relation.
I simplify the notation by omitting the distinction between a Feynman graph and the evaluation of the graph.
One can now twist the antipode with $R$ and define a new map
\bq
 S_R(t) 
 & = & 
 - R \left( t + \sum\limits_{\mathrm{adm. cuts} \; C \; \mathrm{of} \; t} S_R \left( P^C(t) \right) \cdot R^C(t) \right).
\eq
From the multiplicativity constraint (\ref{rotabaxter}) it follows that
\bq
 S_R\left(t_1 t_2 \right)
 & = &
 S_R\left(t_1 \right)
 S_R\left(t_2 \right).
\eq
If we replace $S$ by $S_R$ in (\ref{untwisted})
we obtain
\bq
\label{twisted}
 S_R\left(t^{(1)}\right) t^{(2)}
 & = & 
 \mathrm{finite},
\eq
because by definition $S_R$ differs from $S$ only by finite terms.
Eq. (\ref{twisted}) is equivalent to the forest formula \cite{Zimmermann:1969jj}.
It should be noted that $R$ is not unique and different choices for $R$ correspond
to different renormalization prescriptions.
There is certainly more that could be said on the Hopf algebra of renormalization.
In this regard, we refer the reader to the original 
literature \cite{Krajewski:1998xi,Connes:1998qv,Kreimer:1998iv,Connes:1999yr,Connes:2000fe,vanSuijlekom:2006fk,Ebrahimi-Fard:2010,Ebrahimi-Fard:2012}.

\subsection{Multiple polylogarithms}
\label{subsect:polylogs}

Let us now revisit eq.~(\ref{Feynman_integral_3}) and ask, which functions occur in the coefficients $c_j$.
For one-loop integrals there is a satisfactory answer:
If we restrict our attention to the coefficients $c_j$ with $j\le 0$ (i.e., to $c_{-2}$, $c_{-1}$ and $c_0$), then these coefficients can be expressed
as a sum of algebraic functions of the scalar products of the external momenta and the mass times two transcendental functions,
whose arguments are again algebraic functions of the scalar products and the mass.

The two transcendental functions are the logarithm and the dilogarithm:
\bq
\label{log_dilog}
 \mathrm{Li}_1(x) & = & \sum\limits_{n=1}^\infty \frac{x^n}{n} = - \ln(1-x),
 \nonumber \\
 \mathrm{Li}_2(x) & = & \sum\limits_{n=1}^\infty \frac{x^n}{n^2}.
\eq

\subsubsection{Sum representation of multiple polylogarithms}
\label{sum_repr_polylogs}

Beyond one loop an answer to the above question is not yet known.
However, we know that the following generalizations occur:
From eq.~(\ref{log_dilog}) it is not too difficult to imagine that the generalization includes the classical polylogarithms
defined by
\bq
 \mathrm{Li}_m(x) & = & \sum\limits_{n=1}^\infty \frac{x^n}{n^m}.
\eq
However, explicit calculations for two loops and beyond show that a wider generalization towards functions of several variables is needed and one arrives at the 
multiple polylogarithms defined by \cite{Goncharov_no_note,Goncharov:2001,Borwein}
\bq 
\label{def_multiple_polylogs_sum}
 \mathrm{Li}_{m_1,...,m_k}(x_1,...,x_k)
  & = & \sum\limits_{n_1>n_2>\ldots>n_k>0}^\infty
     \frac{x_1^{n_1}}{{n_1}^{m_1}}\ldots \frac{x_k^{n_k}}{{n_k}^{m_k}}.
\eq
The number $k$ is referred to as the depth of the sum representation of the multiple polylogarithm.
Methods for the numerical evaluation of multiple polylogarithms can be found in \cite{Vollinga:2004sn}. 
The values of the multiple polylogarithms at $x_1=...=x_k=1$ are known as multiple $\zeta$-values:
\index{multiple zeta values}
\bq
\zeta_{m_1,...,m_k} & = & \mathrm{Li}_{m_1,m_2,...,m_k}(1,1,...,1) 
 =
 \sum\limits_{n_1 > n_2 > ... > n_k > 0}^\infty 
 \;\;\;
 \frac{1}{n_1^{m_1}} \cdot ... \cdot \frac{1}{n_k^{m_k}}.
\eq
Important specializations of multiple polylogarithms are the harmonic polylogarithms \cite{Remiddi:1999ew,Gehrmann:2000zt}
\bq
H_{m_1,...,m_k}(x) & = & \mathrm{Li}_{m_1,...,m_k}(x,\underbrace{1,...,1}_{k-1}),
\eq
Further specializations leads to Nielsen's generalized polylogarithms \cite{Nielsen}
\bq
S_{n,p}(x) & = & \mathrm{Li}_{n+1,1,...,1}(x,\underbrace{1,...,1}_{p-1}).
\eq
Although many Feynman integrals evaluate to multiple polylogarithms,
it should be noted that there are Feynman integrals that cannot be expressed in terms of this class of functions.
A prominent example is the two-loop sunrise integral with non-zero internal mass.
Here, elliptic generalizations of multiple polylogarithms occur \cite{Bloch:2013tra,Adams:2014vja,Adams:2015gva}.
These are the focus of current research and beyond the scope of these lectures.

\subsubsection{Integral representation of multiple polylogarithms}
\label{integral_repr_polylogs}

In eq.~(\ref{def_multiple_polylogs_sum}) we have defined multiple polylogarithms through a sum representation.
In addition, multiple polylogarithms have an integral representation. 
To discuss the integral representation it is convenient to introduce 
the following functions for $z_k \neq 0$:
\bq
\label{Gfuncdef}
G(z_1,...,z_k;y) & = &
 \int\limits_0^y \frac{dt_1}{t_1-z_1}
 \int\limits_0^{t_1} \frac{dt_2}{t_2-z_2} ...
 \int\limits_0^{t_{k-1}} \frac{dt_k}{t_k-z_k}.
\eq
The number $k$ is referred to as the depth of the integral representation.
In this definition one variable is redundant due to the following scaling relation:
\bq
\label{G_scaling_relation}
G(z_1,...,z_k;y) & = & G(x z_1, ..., x z_k; x y)
\eq
If one further defines $g(z;y) = 1/(y-z)$, then one has
\bq
\frac{d}{dy} G(z_1,...,z_k;y) & = & g(z_1;y) G(z_2,...,z_k;y)
\eq
and
\bq
\label{Grecursive}
G(z_1,z_2,...,z_k;y) & = & \int\limits_0^y dt \; g(z_1;t) G(z_2,...,z_k;t).
\eq
One can slightly enlarge the set and define $G(0,...,0;y)$ with $k$ zeros for $z_1$ to $z_k$ to be
\bq
\label{trailingzeros}
G(0,...,0;y) & = & \frac{1}{k!} \left( \ln y \right)^k.
\eq
This permits us to allow trailing zeros in the sequence
$(z_1,...,z_k)$ by defining the function $G$ with trailing zeros via eq.~(\ref{Grecursive}) 
and eq.~(\ref{trailingzeros}).
The multiple polylogarithms are related to the functions $G$ by conveniently introducing
the following short-hand notation:
\bq
\label{Gshorthand}
G_{m_1,...,m_k}(z_1,...,z_k;y)
 & = &
 G(\underbrace{0,...,0}_{m_1-1},z_1,...,z_{k-1},\underbrace{0...,0}_{m_k-1},z_k;y)
\eq
Here, all $z_j$ for $j=1,...,k$ are assumed to be non-zero.
One then finds
\bq
\label{Gintrepdef}
\mathrm{Li}_{m_1,...,m_k}(x_1,...,x_k)
& = & (-1)^k 
 G_{m_1,...,m_k}\left( \frac{1}{x_1}, \frac{1}{x_1 x_2}, ..., \frac{1}{x_1...x_k};1 \right).
\eq
The inverse formula reads
\bq
G_{m_1,...,m_k}(z_1,...,z_k;y) & = & 
 (-1)^k \; \mathrm{Li}_{m_1,...,m_k}\left(\frac{y}{z_1}, \frac{z_1}{z_2}, ..., \frac{z_{k-1}}{z_k}\right).
\eq
Eq.~(\ref{Gintrepdef}) together with eq.~(\ref{Gshorthand}) and eq.~(\ref{Gfuncdef})
defines an integral representation for the multiple polylogarithms.
As an example, we obtain from eq.~(\ref{Gintrepdef}) and eq.~(\ref{G_scaling_relation}) 
the integral representation of harmonic polylogarithms:
\bq
H_{m_1,...,m_k}(x) & = & 
 \left(-1\right)^k G_{m_1,...,m_k}\left(1,...,1;x\right).
\eq
The function $G_{m_1,...,m_k}(1,...,1;x)$ is an iterated integral \cite{Chen,Brown:2013qva} in which only the two one-forms
\bq
 \omega_0 = \frac{dt}{t},
 & &
 \omega_1 = \frac{dt}{t-1}
\eq
corresponding to $z=0$ and $z=1$
appear. If one restricts the possible values of $z$ to zero and the $n$-th roots of unity, one arrives at the class of cyclotomic harmonic
polylogarithms \cite{Ablinger:2011te}.

\subsubsection{Notation}
\label{section_notation}

Before we discuss the Hopf algebras associated with multiple polylogarithms 
it is worth explaining to mathematical purists the notation which we will use.
Let us consider a Hopf algebra $H$ and an algebra $A$, together with a map
\bq
 f & : & H \rightarrow A.
\eq
The map $f$ is assumed to be an algebra homomorphism; therefore, for $h_1, h_2 \in H$
\bq
 f\left(h_1 \cdot h_2\right) & = & f\left(h_1\right) \cdot f\left(h_2\right).
\eq
On $H$ we additionally have the dual structures (co-unit $\bar{e}$, co-multiplication $\Delta$) and the antipode $S$.
As $A$ is only assumed to be an algebra, these structures do not exist on $A$.
It is sometimes useful to consider the images of $\Delta(h)$, $\bar{e}(h)$, and $S(h)$ under the map $f$ in $A$.
By abuse of notation we will write
\bq
\label{notation_abuse}
 \Delta f(h),
 \;\;\;
 \bar{e} f(h),
 \;\;\;
 S f(h)
\eq
for
\bq
 \left( f \otimes f \right) \Delta(h),
 \;\;\;
 f\left(\bar{e}(h)\right),
 \;\;\;
 f\left( S(h) \right).
\eq
Eq.~(\ref{notation_abuse}) is merely a handy notation and does not define a Hopf algebra on $A$.

In the examples in the next two subsections, 
$H$ will be either a shuffle algebra or a quasi-shuffle algebra,
$A$ the complex numbers ${\mathbb C}$, and the map $f$ will be given by the evaluation of the functions
$G$ or $\mathrm{Li}$, extended linearly on the vector space of words.

\subsubsection{Shuffle algebra of multiple polylogarithms}
\label{section_shuffle}

Multiple polylogarithms have a rich algebraic structure.  
The representations as iterated integrals and nested sums induce a shuffle algebra and a quasi-shuffle algebra, respectively.
Shuffle and quasi-shuffle algebras are Hopf algebras.
Note that the shuffle algebra of multiple polylogarithms is distinct from the quasi-shuffle algebra of multiple polylogarithms.

\begin{figure}
\begin{center}
\includegraphics[scale=0.8]{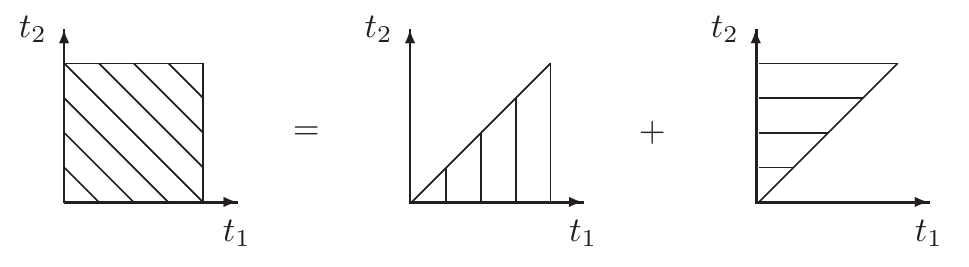}
\caption{\label{fig5} Shuffle algebra from the integral representation: The shuffle product follows from replacing the integral over the square by an integral over the lower triangle and an integral over the upper triangle.}
\end{center}
\end{figure}

We first discuss the shuffle algebra of multiple polylogarithms. The starting point is the integral representation given in eq.~(\ref{Gfuncdef}):
\bq
G(z_1,...,z_k;y) & = &
 \int\limits_0^y \frac{dt_1}{t_1-z_1}
 \int\limits_0^{t_1} \frac{dt_2}{t_2-z_2} ...
 \int\limits_0^{t_{k-1}} \frac{dt_k}{t_k-z_k}.
\eq
For fixed $y$, the ordered sequence of variables $z_1, z_2, ..., z_k$ forms a word $w = z_1 z_2 ... z_k$ and we have the shuffle algebra
\bq
\label{G_shuffle_product}
 G(z_1,z_2,...,z_k;y) \cdot G(z_{k+1},...,z_r; y) = 
 \sum\limits_{\mathrm{shuffles} \; \sigma} G(z_{\sigma(1)},z_{\sigma(2)},...,z_{\sigma(r)};y),
\eq
where the sum includes all permutations $\sigma$, which preserve the relative order
of $1,2,...,k$ and of $k+1,...,r$.
The unit $e$ is given by the empty word:
\bq
 e
 \;\; = \;\;
 G(;y).
\eq
An example for the multiplication is given by
\bq
\label{example_G_product}
G(z_1;y) G(z_2;y) 
 & = & 
 G(z_1,z_2;y) + G(z_2,z_1;y).
\eq
The proof that the integral representation of the multiple polylogarithms fulfills the shuffle product formula in eq.~(\ref{G_shuffle_product})
is sketched for the example in eq.~(\ref{example_G_product}) in fig.~(\ref{fig5})
and can easily be extended to multiple polylogarithms of higher depth by recursively replacing the 
two outermost integrations by integrations over the upper and lower triangle.
For the co-product one has
\bq
 \Delta G(z_1,...,z_k;y) 
 & = & 
 \sum\limits_{j=0}^k G(z_{j+1},...,z_k;y) \otimes G(z_1,...,z_j;y) 
\eq
and for the antipode one finds
\bq
\label{Gantipodeexpli}
 S G(z_1,...,z_k;y) 
 & = & 
 (-1)^k G(z_k,...,z_1;y).
\eq
The shuffle multiplication is commutative; therefore, the antipode satisfies
\bq
 S^2 
 & = & 
 \mathrm{id}.
\eq
From eq.~(\ref{Gantipodeexpli}) this is evident.

\subsubsection{Quasi-shuffle algebra of multiple polylogarithms}
\label{section_quasi_shuffle}

Let us now consider the second Hopf algebra of multiple polylogarithms, which follows from the 
sum representation.
This Hopf algebra is a quasi-shuffle algebra.
A quasi-shuffle algebra is a slight generalization of a shuffle algebra.
Assume that for the set of letters $A$ we have an additional operation
\bq
\label{additional_operation}
 (.,.) & : & A \otimes A \rightarrow A,
 \nonumber \\
       & &  l_1 \otimes l_2 \rightarrow (l_1, l_2),
\eq
which is commutative and associative.
Then we can define a new product of words recursively through
\bq
\label{def_recursive_quasi_shuffle}
\left( l_1 l_2 ... l_k \right) \ast \left( l_{k+1} ... l_r \right) & = &
 l_1 \left[ \left( l_2 ... l_k \right) \ast \left( l_{k+1} ... l_r \right) \right]
+
 l_{k+1} \left[ \left( l_1 l_2 ... l_k \right) \ast \left( l_{k+2} ... l_r \right) \right]
 \nonumber \\
 & &
+
(l_1,l_{k+1}) \left[ \left( l_2 ... l_k \right) \ast \left( l_{k+2} ... l_r \right) \right],
\eq
together with
\bq
 l \ast e 
 \;\; = \;\;
 e \ast l
 \;\; = \;\;
 l.
\eq
This product is a generalization of the shuffle product and differs from the recursive
definition of the shuffle product in eq.~(\ref{def_recursive_shuffle}) through the extra term in the last line of eq.~(\ref{def_recursive_quasi_shuffle}).
This modified product is known under the names quasi-shuffle product \cite{Hoffman},
mixable shuffle product \cite{Guo}, stuffle product \cite{Borwein}, or mould symmetrel \cite{Ecalle}.
Quasi-shuffle algebras are Hopf algebras.
Co-multiplication and co-unit are defined as for the shuffle algebras.
The co-unit $\bar{e}$ is given by:
\bq
\bar{e}\left( e\right) = 1, \;\;\;
& &
\bar{e}\left( l_1 l_2 ... l_n\right) = 0.
\eq
The co-product $\Delta$ is given by:
\bq
\Delta\left( l_1 l_2 ... l_k \right) 
& = & \sum\limits_{j=0}^k \left( l_{j+1} ... l_k \right) \otimes \left( l_1 ... l_j \right).
\eq
The antipode $S$ is recursively defined through
\bq
S\left( l_1 l_2 ... l_k \right) & = & 
 - l_1 l_2 ... l_k
 - \sum\limits_{j=1}^{k-1} S\left( l_{j+1} ... l_k \right) \ast \left( l_1 ... l_j \right),
 \;\;\;\;\;\;
 S(e) = e.
\eq
The sum representation of the multiple polylogarithms in eq.~(\ref{def_multiple_polylogs_sum})
gives rise to a quasi-shuffle algebra.
We determine this by first introducing \cite{Moch:2001zr}
\bq 
 Z(N;m_1,...,m_k;x_1,...,x_k)
  & = & \sum\limits_{N\ge n_1>n_2>\ldots>n_k>0}
     \frac{x_1^{n_1}}{{n_1}^{m_1}}\ldots \frac{x_k^{n_k}}{{n_k}^{m_k}}.
\eq
For $N=\infty$ we recover the multiple polylogarithms:
\bq
 \mathrm{Li}_{m_1,...,m_k}\left(x_1,...,x_k\right)
 & = &
 Z(\infty;m_1,...,m_k;x_1,...,x_k).
\eq
The recursive definition for the quasi-shuffle product of the $Z$-sums reads
\bq
\label{quasi_shuffle_multiplication}
\lefteqn{
 Z(N;m_1,m_2,...,m_k;x_1,x_2,...,x_k) \ast Z(N;m_{k+1},...,m_r;x_{k+1},...,x_r) 
= } & &
 \\
 & &
   \sum\limits_{i_1=1}^{N} \frac{x_1^{i_1}}{i_1^{m_1}} \; Z(i_1-1;m_2,...,m_k;x_2,...,x_k) \ast Z(i_1-1;m_{k+1},...,m_r;x_{k+1},...,x_r)
 \nonumber \\
 & &
 +  \sum\limits_{j_1=1}^{N} \frac{x_{k+1}^{j_1}}{j_1^{m_{k+1}}} \; Z(j_1-1;m_1,...,m_k;x_1,...,x_k) \ast Z(j_1-1;m_{k+2},...,m_k;x_{k+2},...,x_r)
 \nonumber \\
 & &
 +  \sum\limits_{i=1}^{N} \frac{(x_1 x_{k+1})^i}{i^{m_1+m_{k+1}}} \; Z(i-1;m_2,...,m_k;x_2,...,x_k) \ast Z(i-1;m_{k+2},...,m_r;x_{k+2},...,x_r).
 \nonumber
\eq
Note that a letter $l_j$ corresponds to a pair $(m_j;x_j)$.
For $l_1=(m_1;x_1)$ and $l_2=(m_2;x_2)$
the additional operation in eq.~(\ref{additional_operation}) is given by
\bq
 \left(l_1,l_2\right)
 & = &
 \left( m_1+m_2; x_1 x_2 \right).
\eq
A simple example for the quasi-shuffle multiplication is given by
\bq
\label{example_Li_product}
 \mathrm{Li}_{m_1}(x_1) \mathrm{Li}_{m_2}(x_2) 
& = & 
 \mathrm{Li}_{m_1,m_2}(x_1,x_2) + \mathrm{Li}_{m_2,m_1}(x_2,x_1)
                                 + \mathrm{Li}_{m_1+m_2}(x_1x_2).
\eq
The proof that the sum representation of the multiple polylogarithms 
\begin{figure}
\begin{center}
\includegraphics[scale=0.8]{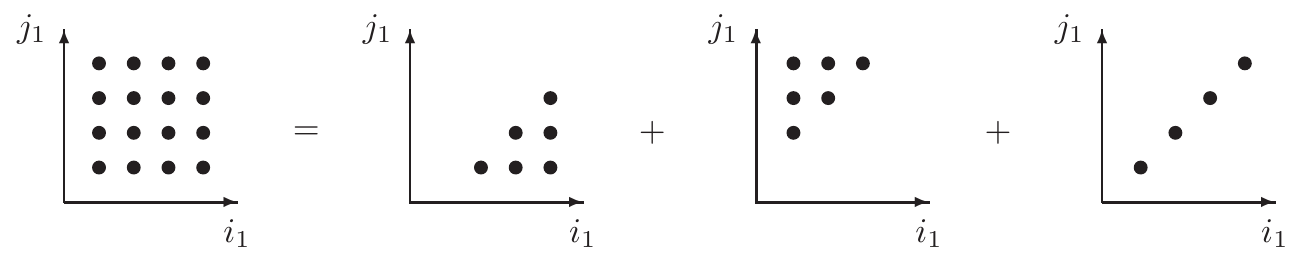}
\caption{\label{fig4} Quasi-shuffle algebra from the sum representation: 
The quasi-shuffle product follows from replacing the sum over the square 
by a sum over the lower triangle, a sum over the upper triangle, and a sum over the diagonal.}
\end{center}
\end{figure}
fulfills the quasi-shuffle product formula in eq.~(\ref{quasi_shuffle_multiplication})
is sketched for the example in eq.~(\ref{example_Li_product}) in fig.~(\ref{fig4})
and can easily be extended to multiple polylogarithms of higher depth by recursively replacing the 
two outermost summations by summations over the upper triangle, the lower triangle, and the diagonal.

Let us provide one further example for the quasi-shuffle product.
Working out the recursive definition of the quasi-shuffle product we obtain
\bq
\lefteqn{
\mathrm{Li}_{m_1,m_2}(x_1,x_2) \cdot \mathrm{Li}_{m_3}(x_3) = } \nonumber \\
& = &  \mathrm{Li}_{m_1,m_2,m_3}(x_1,x_2,x_3) 
+ \mathrm{Li}_{m_1,m_3,m_2}(x_1,x_3,x_2) 
+ \mathrm{Li}_{m_3,m_1,m_2}(x_3,x_1,x_2) 
\nonumber \\ & & 
+ \mathrm{Li}_{m_1,m_2+m_3}(x_1,x_2x_3) 
+ \mathrm{Li}_{m_1+m_3,m_2}(x_1 x_3,x_2) 
\eq
\begin{figure}
\begin{center}
\includegraphics[scale=0.8]{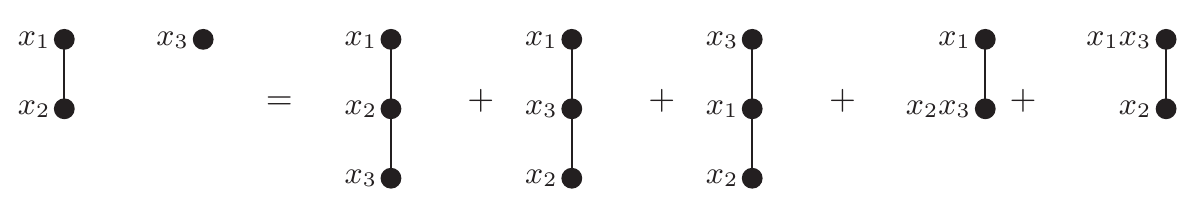}
\caption{\label{fig3} Pictorial representation of the quasi-shuffle multiplication law. The first three terms
on the right-hand side correspond to the ordinary shuffle product, whereas the two last terms are the additional ``stuffle''-terms.}
\end{center}
\end{figure}
This is shown pictorially in fig.~(\ref{fig3}).
The first three terms correspond to the ordinary shuffle product, whereas the two last terms are the additional ``stuffle''-terms.
In fig.~(\ref{fig3}) we show only the $x$-variables, which are multiplied in the stuffle-terms.
Not shown in fig.~(\ref{fig3}) are the indices $m_j$, which are added in the stuffle-terms.

\subsubsection{Hopf algebra related to the Hodge structure}
\label{section_hodge}

The multiple polylogarithms are periods of a mixed Hodge-Tate structure.
From this Hodge structure one obtains a third Hopf algebra as follows \cite{Goncharov:2001,Goncharov:2002b}:
Let $S$ be a set of pairwise distinct points in ${\mathbb C}$.
We denote
\bq
\label{def_I_iterated_integral}
 I\left(z_0;z_1,z_2,...,z_k;z_{k+1}\right)
 & = &
 \int\limits_{z_0}^{z_{k+1}} \frac{dt_k}{t_k-z_k}
 \int\limits_{z_0}^{t_k} \frac{dt_{k-1}}{t_{k-1}-z_{k-1}}
 ...
 \int\limits_{z_0}^{t_2} \frac{dt_1}{t_1-z_1},
\eq
together with the convention
\bq
\label{def_I_unit}
 I\left(z_0;z_1\right) & = & 1.
\eq
This is a slight generalization of eq.~(\ref{Gfuncdef}), allowing the starting point $z_0$ 
of the integration to be different from zero.
We have
\bq
 G\left(z_1,...,z_k;y\right) 
 & = & 
 I\left(0;z_k,...,z_1;y\right),
 \nonumber \\
 I\left(z_0;z_1,z_2,...,z_k;z_{k+1}\right)
 & = &
 G\left(z_k-z_0,...,z_1-z_0;z_{k+1}-z_0\right).
\eq
As an algebra one now considers the ${\mathbb Q}$-algebra generated by the iterated integrals of the form
in eq.~(\ref{def_I_iterated_integral}) together with the relation~(\ref{def_I_unit}).
The expression ``generated by'' means that there are no further relations implied, and a product such as
\bq
 I\left(z_0;z_1;z_2\right) \cdot I\left(z_3;z_4;z_5\right)
\eq
is left as it is.
The co-product is more interesting.
We define it by treating the quantities
$I(z_0;z_1,...,z_k;z_{k+1})$ 
as abstract objects and we set
\bq
\lefteqn{
 \Delta
 I\left(z_0;z_1,z_2,...,z_k;z_{k+1}\right)
 = 
 \sum\limits_{r=0}^k
 \;\;
 \sum\limits_{0 = i_0 < i_1 < ... < i_r < i_{r+1} = k+1}
} & & 
 \nonumber \\
 & &
 \prod\limits_{p=0}^r
 I\left(z_{i_p};z_{i_p+1},z_{i_p+2},...,z_{i_{p+1}-1};z_{i_{p+1}}\right)
 \otimes
 I\left(z_0;z_{i_1},z_{i_2},...,z_{i_r};z_{k+1}\right).
\eq
In \cite{Goncharov:2002b} it is shown, that this defines a Hopf algebra.
We remind the reader that we use a sloppy notation, as explained in section~\ref{section_notation}.
For rigorous mathematicians the co-product is defined on the algebra of strings of the form
$(z_0;z_1,...,z_k;z_{k+1})$.
Furthermore, in addition one can consider this Hopf algebra modulo by considering the following relations:
For identical start and end points of the integration one can impose 
the shuffle relation:
\bq
\label{I_shuffle}
 I(z_0;z_1,...,z_k; z_{r+1}) \cdot I(z_0;z_{k+1},...,z_r; z_{r+1}) 
 & = &
 \sum\limits_{\mathrm{shuffles} \; \sigma} I(z_0; z_{\sigma(1)},...,z_{\sigma(r)}; z_{r+1}).
 \nonumber \\
\eq
The second relation is the path composition formula:
\bq
 I(z_0;z_1,...,z_r;z_{r+1})
 & = &
 \sum\limits_{k=0}^r
 I(z_0;z_1,...,z_k; y) \cdot I(y;z_{k+1},...,z_r; z_{r+1}).
\eq
Finally, one sets for $k \ge 1$
\bq
\label{I_zero}
 I\left(z_0;z_1,...,z_k;z_0\right) & = & 0.
\eq
Imposing the relations in eqs.~(\ref{I_shuffle})-(\ref{I_zero}) still produces a Hopf algebra.
Let us emphasize that in eq.~(\ref{def_I_iterated_integral}) we assume the points $z_1$, ..., $z_k$ to be pairwise
distinct and each point not equal to $z_0$ nor to $z_{k+1}$.
If this condition is not met, we might have to deal with divergent integrals.
This point is discussed in detail in the original literature \cite{Goncharov:2001,Goncharov:2002b} 
and the lectures by C. Duhr \cite{Duhr:2014woa}.

\subsubsection{Comparison of the various coproducts}

We have now seen three different co-products for the pre-images of the multiple polylogarithms.
We recall that the multiple polylogarithms can be viewed 
as a map from the shuffle algebra to ${\mathbb C}$ (discussed in section~\ref{section_shuffle}), 
as a map from the quasi-shuffle algebra to ${\mathbb C}$ (discussed in section~\ref{section_quasi_shuffle})
or as map from the algebra of strings of the form $(z_0;z_1,...,z_k;z_{k+1})$ to ${\mathbb C}$
(discussed in section~\ref{section_hodge}).
In all three cases we have a co-product on the domain of the map (but not on the co-domain ${\mathbb C}$).
We remind the reader of our notation introduced in section~\ref{section_notation}.
It is instructive to discuss the differences between the various co-products.
We consider the example
\bq
 G\left(z_1,z_2;1\right)
 \;\; = \;\;
 \mathrm{Li}_{11}\left(\frac{1}{z_1},\frac{z_1}{z_2}\right)
 \;\; = \;\;
 I\left(0;z_2,z_1;1\right).
\eq
For the shuffle algebra we have
\bq
 \Delta^{\mathrm{shuffle}} G\left(z_1,z_2;1\right)
 & = &
 G\left(z_1,z_2;1\right) \otimes e + e \otimes G\left(z_1,z_2;1\right)
 \nonumber \\
 & &
 + G\left(z_2;1\right) \otimes G\left(z_1;1\right).
\eq
For the quasi-shuffle algebra we obtain
\bq
 \Delta^{\mathrm{quasi-shuffle}} \mathrm{Li}_{11}\left(\frac{1}{z_1},\frac{z_1}{z_2}\right)
 & = &
 \mathrm{Li}_{11}\left(\frac{1}{z_1},\frac{z_1}{z_2}\right) \otimes e + e \otimes \mathrm{Li}_{11}\left(\frac{1}{z_1},\frac{z_1}{z_2}\right)
 \nonumber \\
 & & 
 + \mathrm{Li}_{11}\left(\frac{z_1}{z_2}\right) \otimes \mathrm{Li}_{11}\left(\frac{1}{z_1}\right).
\eq
Translated to the $G$-notation this reads
\bq
 \Delta^{\mathrm{quasi-shuffle}} G\left(z_1,z_2;1\right)
 & = &
 G\left(z_1,z_2;1\right) \otimes e + e \otimes G\left(z_1,z_2;1\right)
 \nonumber \\
 & &
 + G\left(\frac{z_2}{z_1};1\right) \otimes G\left(z_1;1\right).
\eq
Finally, for the Hopf algebra related to the Hodge structure we obtain
\bq
 \Delta^{\mathrm{Hodge}} I\left(0;z_2,z_1;1\right)
 & = &
 I\left(0;z_2,z_1;1\right) \otimes e + e \otimes I\left(0;z_2,z_1;1\right)
 \\
 & &
 + I\left(0;z_2;z_1\right) \otimes I\left(0;z_1;1\right)
 + I\left(z_2;z_1;1\right) \otimes I\left(0;z_2;1\right).
 \nonumber 
\eq
Again, translating to the $G$-notation we find
\bq
 \Delta^{\mathrm{Hodge}} G\left(z_1,z_2;1\right)
 & = &
 G\left(z_1,z_2;1\right) \otimes e + e \otimes G\left(z_1,z_2;1\right)
 \\
 & &
 + G\left(\frac{z_2}{z_1};1\right) \otimes G\left(z_1;1\right)
 + G\left(\frac{z_1-z_2}{1-z_2};1\right) \otimes G\left(z_2;1\right).
 \nonumber
\eq
We see that the three co-products are different.


\section{Dyson-Schwinger equations}
\label{sec:dyson_schwinger}

We now turn our attention to Dyson-Schwinger equations.
One of the fundamental concepts of quantum field theory is Green's functions.
For a specified set of external particles, Green's function can be thought of as the set of all
Feynman diagrams (to all loop orders) with the specified external particles.
It will be convenient to represent the set of all possible Feynman diagrams for a given set of external
particles by a blob.
\begin{figure}
\begin{center}
\includegraphics[scale=0.7]{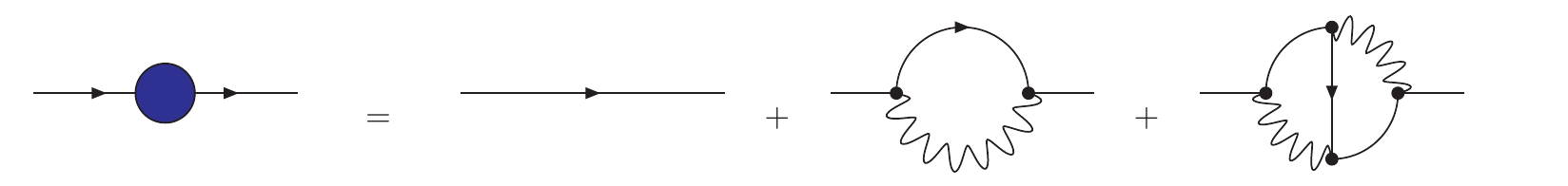}
\includegraphics[scale=0.7]{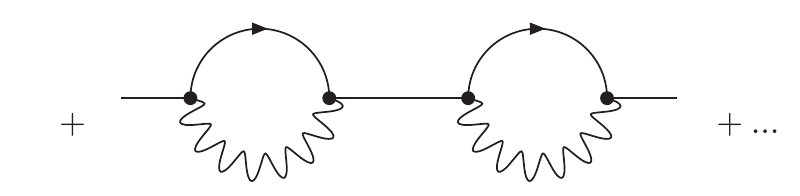}
\includegraphics[scale=0.7]{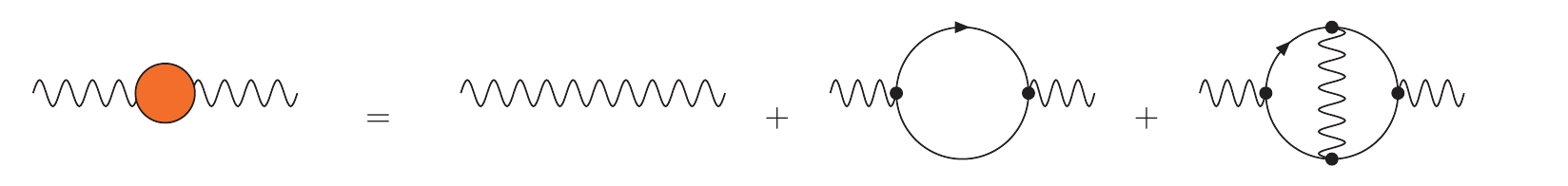}
\includegraphics[scale=0.7]{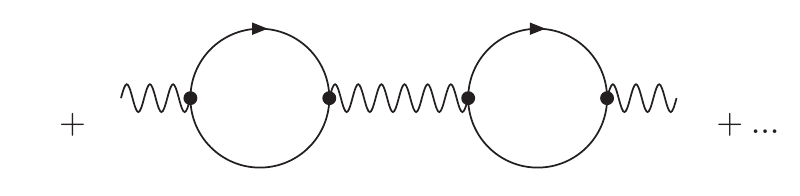}
\includegraphics[scale=0.7]{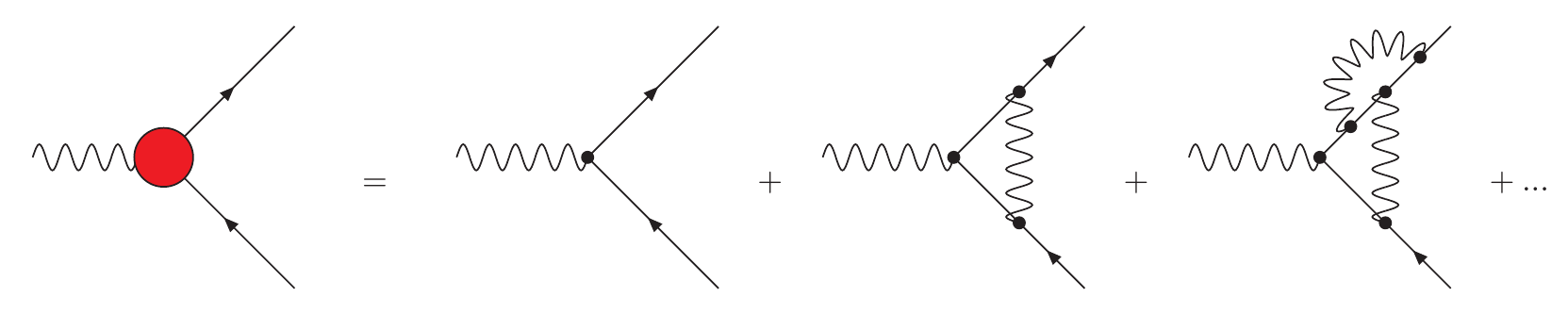}
\caption{\label{fig6} Two-point and three-point functions in QED. 
In the two-point case the blob represents all possible Feynman diagrams with the specified set of external particles,
in the three-point case the blob represent all possible one-particle irreducible Feynman diagrams with the specified set of external particles.
}
\end{center}
\end{figure}
In fig.~(\ref{fig6}) this is illustrated for the two-point and three-point functions in quantum electrodynamics (QED).
In QED we have as two-point function the electron propagator and the photon propagator.
As three-point function we have the electron-photon-vertex function.
The Dyson-Schwinger equations are integral equations among Green's functions.
As an example, the Dyson-Schwinger equations for the electron propagator and the photon propagator in QED
\begin{figure}
\begin{center}
\includegraphics[scale=0.8]{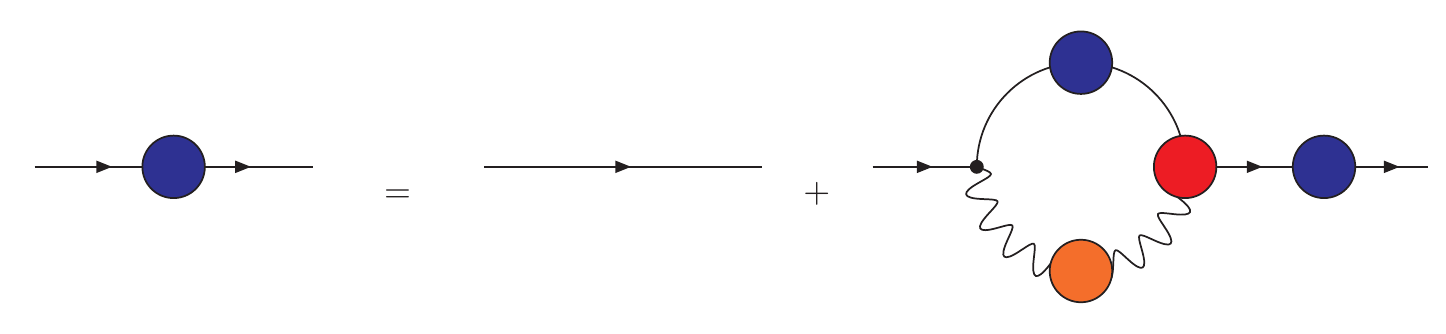}
\includegraphics[scale=0.8]{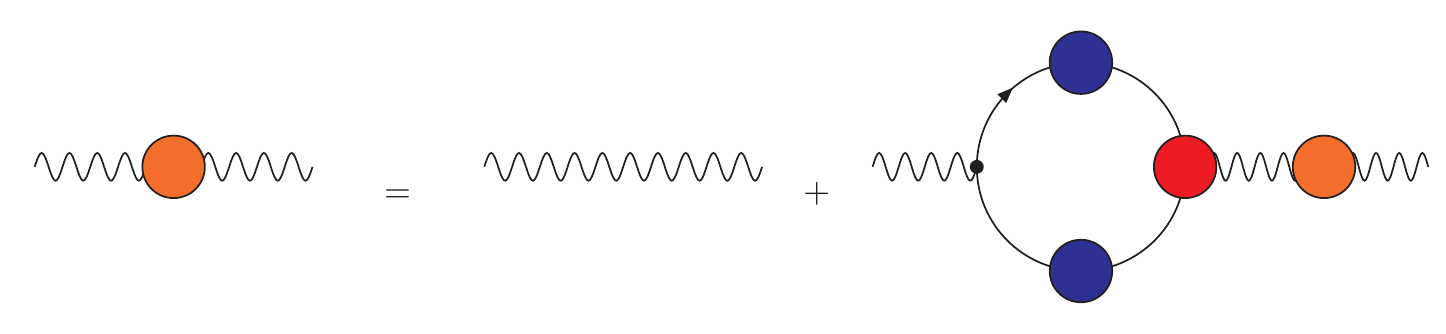}
\caption{\label{fig7} Dyson-Schwinger equations for the electron propagator and the photon propagator in QED.}
\end{center}
\end{figure}
are shown in fig.~(\ref{fig7}).
Note that the Dyson-Schwinger equations for the propagators involve Green's function for the vertex.
In other words, a Dyson-Schwinger equation for a two-point function involves a three-point function.
Let us then look at the Dyson-Schwinger equation for the electron-photon vertex.
\begin{figure}
\begin{center}
\includegraphics[scale=0.8]{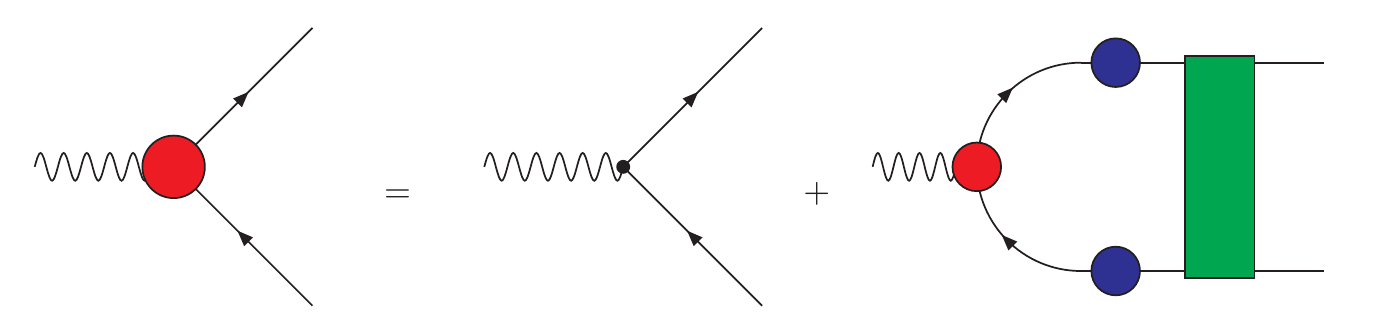}
\caption{\label{fig8} Dyson-Schwinger equation for the electron-photon vertex.}
\end{center}
\end{figure}
This Dyson-Schwinger equation is shown in fig.~(\ref{fig8}) and involves the electron-positron scattering kernel,
depending on four external particles.
This leads to a coupled system of Dyson-Schwinger equations for Green's functions involving all possible 
numbers of external particles.
In order to solve a Dyson-Schwinger equation we have to truncate the system.
We discuss this for a simple example.
Consider a toy model consisting of a fermion and a scalar particle.
We perform two simplifications: First, we linearize the Dyson-Schwinger equation.
In our toy model this implies that the Dyson-Schwinger equation for the scalar-fermion vertex reduces
\begin{figure}
\begin{center}
\includegraphics[scale=0.8]{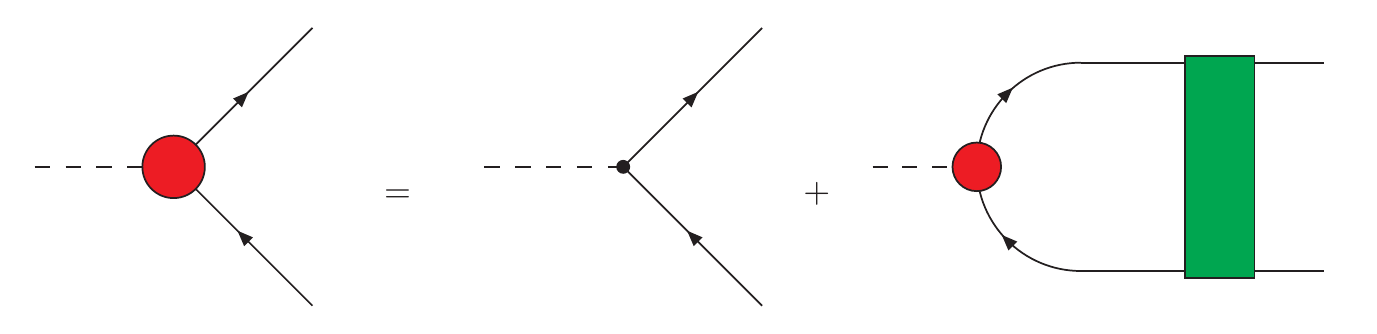}
\caption{\label{fig9} Linearised Dyson-Schwinger equation for a vertex}
\end{center}
\end{figure}
to that shown in fig.~(\ref{fig9}).
In comparison with fig.~(\ref{fig8}) we have replaced the full fermion propagator (the two blue blobs) with
the corresponding Born propagator.
In this way, the unknown function (the scalar-fermion vertex function) appears linearly on the right-hand side
(not multiplied by any other unknown function).
Secondly, we truncate the kernel at a certain loop order.
\begin{figure}
\begin{center}
\includegraphics[scale=0.8]{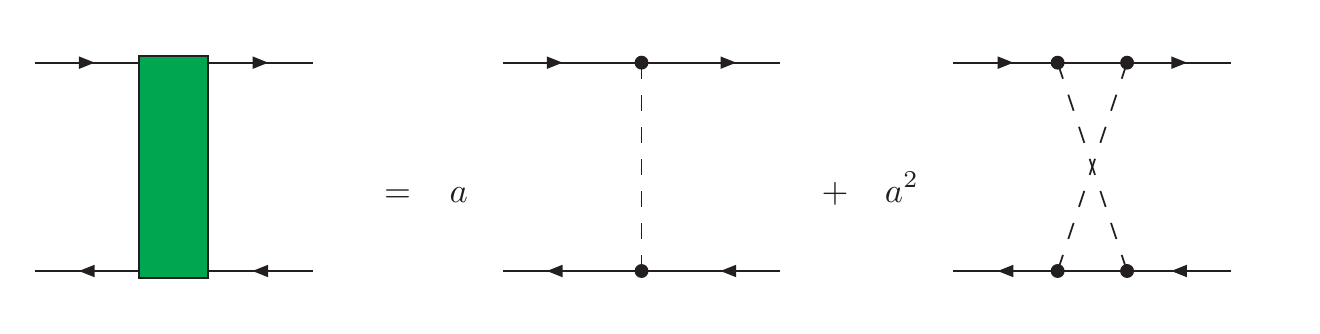}
\caption{\label{fig10} Truncation of the kernel at two-loop order.}
\end{center}
\end{figure}
For example, we could truncate the kernel at two-loop order, shown in fig.~(\ref{fig10}).
The coupling constant is denoted by $a$.
After this truncation, the kernel can be considered a known function.
In other words, with a truncated kernel the Dyson-Schwinger equation in eq.~(\ref{fig9})
is a linear integral equation for the unknown scalar-fermion vertex (the red blob).
Let us make one further simplification by setting the momentum of the scalar external particle to zero, as shown in fig.~(\ref{fig17})
and let us consider massless particles.
\begin{figure}
\begin{center}
\includegraphics[scale=0.8]{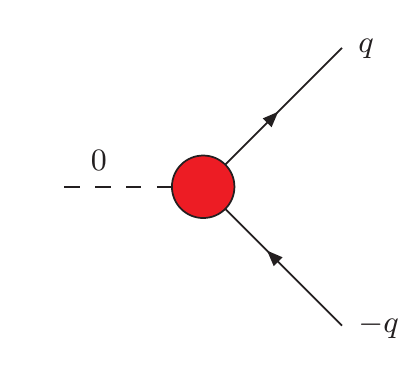}
\caption{\label{fig17} Simplified kinematics for the vertex function: The external scalar particle has zero momentum.
The fermion and the anti-fermion have (outgoing) momentum $q$ and $(-q)$, respectively.}
\end{center}
\end{figure}
The renormalized vertex function $G_R(a,L)$ then only depends on the coupling $a$ and the quantity
\bq
 L & = & \ln\left(\frac{-q^2}{\mu^2}\right),
\eq
where $\mu$ is an arbitrary scale.
As renormalization condition we impose
\bq
\label{renorm_condition}
 G_R\left(a,0\right)
 & = & 
 1.
\eq
From dimensional analysis it follows that $G_R(a,L)$ must be of the form
\bq
\label{functional_form_G_R}
 G_R\left(a,L\right)
 & = & 
 \exp\left(-\gamma_G\left(a\right) L \right).
\eq
The anomalous dimension $\gamma_G$ depends only on the coupling $a$, but not on $L$.
Plugging eq.~(\ref{functional_form_G_R}) into the truncated and linearized Dyson-Schwinger equation
one obtains
\bq
\label{Dyson_Schwinger_example}
\exp\left(-\gamma_G(a)L \right)
 & = &
 1
 +
 \left( \exp\left(-\gamma_G(a)L\right) -1\right)\left[aF_1(\gamma_G)+a^2F_2(\gamma_G)\right],
\eq
where $F_1$ and $F_2$ are the Mellin-transforms of the one-loop and two-loop integral, respectively.
Working these out, one finds
\bq
\label{eq_gamma_G_1}
 1
 & = & 
 -a\frac{1}{\gamma_G(1-\gamma_G)}
 \\
 & &
 -a^2
  \left\{ 
         \frac{1}{\gamma_G^2(1-\gamma_G)^2}
         -4\sum_{n=1}^\infty n(1-2^{-2n})\zeta_{2n+1}\left[\gamma_G^{2n-2}+(1-\gamma_G)^{2n-2}\right]
  \right\}.
 \nonumber
\eq
The sum can be obtained:
\bq
\label{eq_gamma_G_2}
\lefteqn{
 -4\sum_{n=1}^\infty n(1-2^{-2n})\zeta_{2n+1}\left[\gamma_G^{2n-2}+(1-\gamma_G)^{2n-2}\right]
 = } & &
 \\
 & = &
      \frac{1}{\gamma_G} \left[ \psi'\left(1+\gamma_G\right) - \psi'\left(1-\gamma_G\right) \right]
    + \frac{1}{1-\gamma_G} \left[ \psi'\left(2-\gamma_G\right) - \psi'\left(\gamma_G\right) \right]
 \nonumber \\
 & & 
    - \frac{1}{2\gamma_G} \left[ \psi'\left(1+\frac{\gamma_G}{2}\right) - \psi'\left(1-\frac{\gamma_G}{2}\right) \right]
 \nonumber \\
 & & 
    - \frac{1}{2(1-\gamma_G)} \left[ \psi'\left(\frac{3-\gamma_G}{2}\right) - \psi'\left(\frac{1+\gamma_G}{2}\right) \right].
 \nonumber
\eq
Eq.~(\ref{eq_gamma_G_1}) and eq.~(\ref{eq_gamma_G_1}) implicitly define $\gamma_G$ as a function of $a$.
Given $a$, we may solve numerically for $\gamma_G$ \cite{Bierenbaum:2006gn}.

Let us now consider the Hopf algebra side of this example.
\begin{figure}
\begin{center}
\includegraphics[scale=0.7]{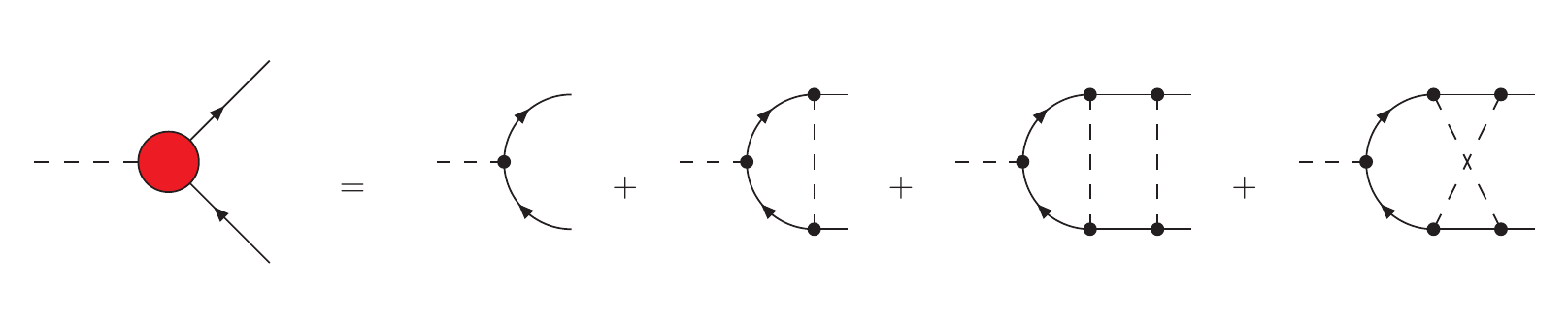}
\includegraphics[scale=0.7]{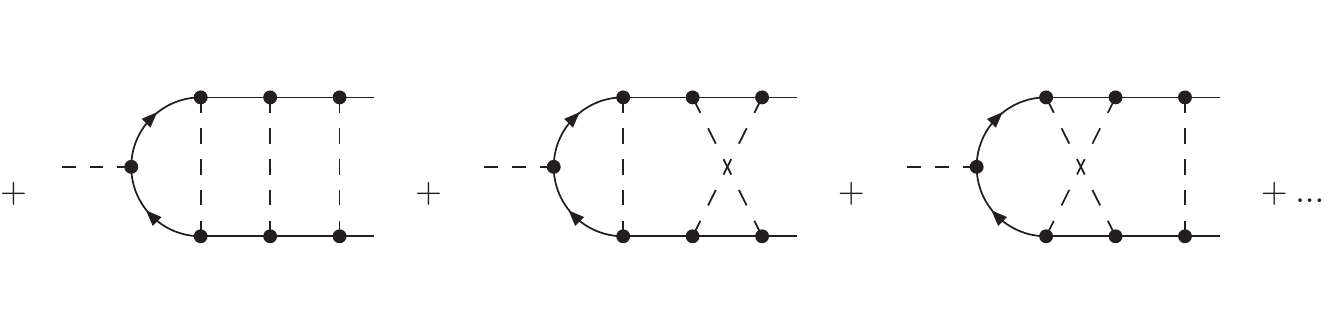}
\caption{\label{fig11} Feynman diagrams computed by the truncated and linearized Dyson-Schwinger equation.}
\end{center}
\end{figure}
Fig.~(\ref{fig11}) shows some Feynman diagrams, which are computed by the truncated and linearized Dyson-Schwinger
equation.
\begin{figure}
\begin{center}
\includegraphics[scale=0.8]{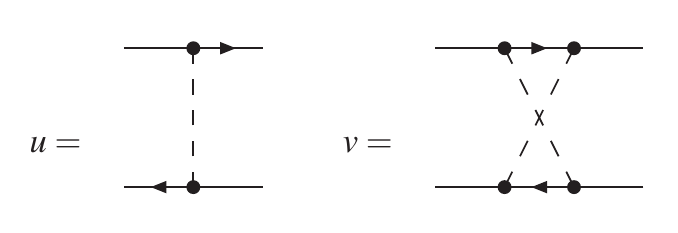}
\caption{\label{fig12} Two letters, corresponding to the one-loop and two-loop contribution to the truncated kernel, respectively.}
\end{center}
\end{figure}
It is convenient, to introduce two letters $u$ and $v$, as shown in fig.~(\ref{fig12}).
The two letters correspond to the one-loop and two-loop contribution to the truncated kernel in fig.~(\ref{fig10}).
With the help of these two letters, we may represent each Feynman diagram in fig.~(\ref{fig11}) by a word
\begin{figure}
\begin{center}
\includegraphics[scale=0.8]{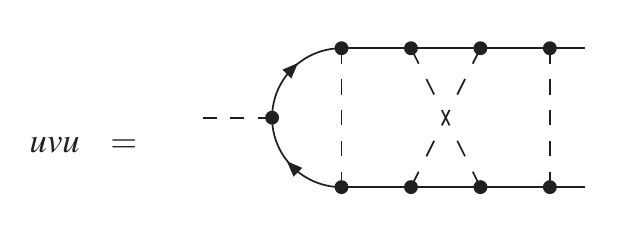}
\caption{\label{fig13} Feynman diagram from fig.~(\ref{fig11}) as represented by a word in the two letters $u$ and $v$.}
\end{center}
\end{figure}
in these two letters, an example is shown in fig.~(\ref{fig13}).
Let us further define two insertion operators 
$B_+^u$ and $B_+^v$ by the action on any words $w$:
\bq 
 B_+^u w = u w,
 & &
 B_+^v w = v w.
\eq
With the help of these two operators we may re-write the Dyson-Schwinger equation in
fig.~(\ref{fig9}) as 
\bq
\label{combinatorial_Dyson_Schwinger}
 X(a) & = & 
 1 + a B_+^u X(a) + a^2 B_+^v X(a)
\eq
Eq.~(\ref{combinatorial_Dyson_Schwinger}) is known as a combinatorial Dyson-Schwinger equation.
In comparing the combinatorial Dyson-Schwinger 
eq.~(\ref{combinatorial_Dyson_Schwinger}) with eq.~(\ref{Dyson_Schwinger_example})
we see that inserting the letter $u$ corresponds in Mellin space
to the multiplication of the Mellin-transform of the one-loop integral with the subtracted Green's function.
The subtraction for Green's function implements ultraviolet renormalization
and the renormalization condition in eq.~(\ref{renorm_condition}).
In a similar way, the letter $v$ corresponds 
to the multiplication of the Mellin-transform of the two-loop integral with the subtracted Green's function.

A solution to eq.~(\ref{combinatorial_Dyson_Schwinger}) is given by
\bq
 X(a)
 & = &
 \exp_\Sha\left( a u + a^2 v \right),
\eq
where $\exp_\Sha$ denotes the shuffle-exponential
\bq
 \exp_\Sha\left(w\right)
 & = &
 \sum\limits_{n=0}^\infty \frac{1}{n!} w^{\Sha n}
\eq
and $\Sha$ denotes the shuffle product:
\bq
 u^{\Sha n} & = & \underbrace{\; u \; {\scriptstyle \Sha} \; u \; {\scriptstyle \Sha} \; ... \; {\scriptstyle \Sha} \; u \;}_{n} 
 = n! \; \underbrace{\; u u ... u \;}_{n},
 \nonumber \\
 u {\scriptstyle \Sha} v & = & u v + v u.
\eq
In terms of Hopf algebra we obtain the shuffle algebra in the two letters $u$ and $v$.
For the first few terms of $X(a)$ in an expansion in $a$ we have
\bq
 X(a)
 & = &
 1 + a u + a^2 \left( u u + v \right) + a^3 \left( u u u + u v + v u \right) + ...
\eq
$X(a)$ is a group-like element in this Hopf algebra. Therefore, for the co-product we have
\bq
 \Delta X(a) & = &
 X(a) \otimes X(a).
\eq
Combinatorial Hopf algebras are currently the subject of studies \cite{Bergbauer:2005fb,Kreimer:2006gm,Foissy:2011,Krueger:2014poa}.
Although we only discussed a simple example here, more complicated cases can be envisaged.
The challenge is to map the iterated structure of a Feynman graph to the iterated structure
of the functions to which this graph evaluates.
Techniques such as Mellin-Barnes \cite{Bierenbaum:2003ud,Kreimer:2012nk,Panzer:2014kia},
linear reducibility \cite{Brown:2008}, and algorithms based on
nested sums \cite{Moch:2001zr} may prove useful in this respect.
Examples of recent research in the field of Dyson-Schwinger equations 
can be found in \cite{Broadhurst:2000dq,Kreimer:2006ua,vanBaalen:2008tc,vanBaalen:2009hu,Bellon:2013sya,Bellon:2014,Clavier:2014osa}.


\end{document}